%% file: Chaos2018.tex
\documentclass[aip,reprint,cha]{revtex4-1}   	
\usepackage[pdftex]{graphicx}

\usepackage[usenames,dvipsnames,svgnames,table]{xcolor}
\definecolor{black}{rgb}{0,0,0}
\definecolor{myblue}{rgb}{0,0,0.6}
\definecolor{mygreen}{rgb}{0,0.5,0}

\usepackage{amsmath,amssymb}
\usepackage{comment}
\usepackage{bbold}
\usepackage[normalem]{ulem}
\usepackage{tabularx}
\usepackage{multirow}
\usepackage{subfigure}

\newif\ifcolor
\colortrue

\begin{document}

\title{Robust Approach for Rotor Mapping in Cardiac Tissue}
\author{Daniel R. Gurevich}
\affiliation{School of Physics, Georgia Institute of Technology, Atlanta, GA 30332, USA}
\author{Roman O. Grigoriev}
\affiliation{School of Physics, Georgia Institute of Technology, Atlanta, GA 30332, USA}
\date{\today}						

\begin{abstract}
The motion of and interaction between phase singularities that lie at the centers of spiral waves captures many qualitative and, in some cases, quantitative features of complex dynamics in excitable systems. 
Being able to accurately reconstruct their position is thus quite important, even if the data are noisy and sparse, as in electrophysiology studies of cardiac arrhythmias, for instance. 
A recently proposed global topological approach [Marcotte \& Grigoriev, Chaos 27, 093936 (2017)] promises to meaningfully improve the quality of the reconstruction compared with traditional, local approaches. 
Indeed, we found that this approach is capable of handling noise levels exceeding the range of the signal with minimal loss of accuracy. 
Moreover, it also works successfully with data sampled on sparse grids with spacing comparable to the mean separation between the phase singularities for complex patterns featuring multiple interacting spiral waves.
\end{abstract}

\pacs{}

\keywords{phase singularity, cardiac arrhythmia, spiral wave chaos}

\maketitle

\begin{quotation}
Catheter ablation has recently emerged as a leading medical treatment for a range of cardiac arrhythmias, especially atrial fibrillation. 
The premise of the treatment is that certain localized regions of heart tissue can become sources of spiral excitation waves -- or rotors -- competing with the heart's natural pacemaker, i.e., the sinoatrial node for the atria or the atrio-ventricular node for the ventricles. 
The success of the ablation procedure then critically depends on the precision with which these sources are located based on electrograms obtained using intra-cardiac multi-electrode catheters. 
This paper explains how the sources of excitation waves in a numerical model of atrial fibrillation can be reliably located with subgrid precision using sparse and noisy measurements of the transmembrane voltage.
A similar approach could be used to improve the quality of rotor mapping in a clinical setting.
\end{quotation}

\section{Introduction}

Spiral waves in two dimensions (and scroll waves in three dimensions) represent the key motifs of typical self-sustained dynamical patterns in excitable systems such as cardiac tissue.
In fact, the work to understand the mechanisms and develop effective treatments of cardiac arrhythmias such as tachycardia or fibrillation became the major driver for much of the recent interest in the dynamics of spiral and scroll waves. 
Because of the strong spatial coherence of such waves, many aspects of their dynamics can in fact be understood using center manifold reduction of the underlying partial differential (or difference) equations, yielding a system of ordinary differential equations with respect to just a few variables \cite{Barkley94,sandstede1997}.
These variables are associated with the Euclidean symmetry of the problem and can be interpreted in terms of the low-dimensional dynamics (translation and rotation) of the core of the spiral wave, which serves as its source and anchor.
Notable examples include meander and drift \cite{Biktashev:1995re,BiHoNi96,FiSaScWu96,FiTu98} of spiral waves and their interaction with boundaries \cite{LanBar13,LanBar14,Marcotte2016}.
In fact, even for complex patterns of excitation which involve multiple spiral waves, many features of the dynamics can be understood and described reasonably well in terms of the wave core interaction \cite{ByMaGr14,Marcotte2016}.

Given their influence on large regions of space, spiral wave cores represent attractive targets for controlling the dynamics of excitable media.
This is well-known to clinical practitioners looking for treatments of cardiac arrhythmias such as atrial fibrillation.
In fact, radio frequency ablation, which aims to silence or isolate regions of cardiac tissue believed to be sources of spiral excitation waves (often referred to as rotors), has become the leading surgical treatment for persistent atrial fibrillation \cite{narayan2012,shivkumar2012}.
The success rate of ablation surgeries is however not very high, suggesting that the insufficient accuracy with which the spiral wave cores are located may be problematic.
Indeed, typical intra-cardiac basket catheters used to locate them only have 64 unipolar electrodes distributed over 8 circular spines of the catheter \cite{laughner2016}.
The signal they generate is both rather sparse and rather noisy, making it challenging to identify the location of the spiral wave cores, especially if those cores are drifting or meandering.

This paper describes a novel method that can reliably identify and track with high precision a large number of spiral wave cores based on sparse and noisy measurements of the transmembrane voltage.
The paper is organized as follows.
An overview of the existing methods for identifying the cores (or, more precisely, certain points inside the cores) is provided in Section \ref{sec:background}. 
Our approach is described in Section \ref{sec:methods}, and it is validated and compared with competing approaches in Section \ref{sec:results}.
The limitations and potential extensions of our approach are discussed in Section \ref{sec:discussion}.
Finally, our conclusions are presented in Section \ref{sec:conclusions}.

\section{Background}
\label{sec:background}

Spiral wave cores (defined in terms of the response functions that determine the sensitivity to perturbations, heterogeneity, etc.) tend to be exponentially localized, as illustrated by analyses of the Ginzburg-Landau \cite{biktasheva2001}, Barkley \cite{Henry2002}, FitzHugh-Nagumo \cite{BiHoBi06}, Oregonator \cite{biktasheva2015}, Beeler-Reuter-Pumir \cite{Biktashev2011}, and Karma \cite{Marcotte2016} models.
In practice, it is more convenient to define a single point that characterizes the position of the core.
Numerical studies tend to use the position of the spiral tip, which can be defined in many different ways \cite{Fenton2002, biktashev1994tension, zhang1995chaotic, BaKnTu90, jahnke1989chemical, Fenton1998, biktashev1994tension, henze1990helical, ByMaGr14, FouBik10, bar1994spiral, beaumont1998spiral, jahnke1989chemical, jahnke1991survey}. 
The most popular definitions are based on either the intersection of level sets of different variables \cite{BaKnTu90} or the vanishing of the normal velocity \cite{Fenton1998} or the curvature of the wavefront \cite{beaumont1998spiral}.

Neither of these definitions are convenient (or reliable) for analyzing experimental data, however. 
An alternative approach relies on the phase-amplitude representation of spiral waves, with the phase singularity (PS) defining the instantaneous center of rotation of the wave.
A method for identifying PSs based on the local phase field has become standard in analyzing experimental data \cite{Gray1997,iyer2001experimentalist}, although it is also possible to determine the location of PSs using the amplitude field \cite{ByMaGr14}.

Recently, several alternative approaches have been proposed to locate spiral wave cores based on various metrics such as Shannon entropy, multi-scale frequency, kurtosis, multi-scale entropy \cite{annoni2018}, and Jacobian determinant \cite{li2018}.
It should be noted that these metrics define neither the spiral tip nor the phase singularity, but they can be applied to sparse data, although their precision and accuracy decrease very quickly as sparsity increases.  
The study of Li {\it et al.} \cite{li2018} showed that the Jacobian determinant method locates  both stationary and meandering spirals with precision higher than other common approaches\cite{iyer2001experimentalist, bray2001, lee2016new, Fenton1998}.
Furthermore, they determined that the Jacobian determinant method is the only one that can produce reliable results in the presence of as much as 0.9\% noise.

In fact, essentially all existing methods for locating spiral wave cores are local and cannot withstand higher noise levels characteristic of practical applications without significant loss of accuracy and precision.
The only exception is the global topological method for identifying PSs proposed by Marcotte and Grigoriev \cite{marcottetop}.
The original version of the topological approach defined PSs as intersections of level sets associated with two different variables (one fast, one slow).
A modified version of this topological approach based on phase reconstruction that required measurement of just one variable was developed and tested using spatially resolved numerical and experimental data \cite{gurevich2017level}.
Here we describe a robust implementation of the topological approach that does not require phase reconstruction and investigate its performance for sparse and noisy data generated by a model of atrial fibrillation.
In addition to its relative simplicity, the implementation described here also affords a more direct dynamical interpretation and allows an automatic classification of topological changes leading to creation or destruction of pairs of counterrotating rotors \cite{marcottetop}.

\section{Methods}
\label{sec:methods}

\subsection{Model}

To illustrate the algorithm and determine the conditions under which it functions reliably, we will use two-dimensional surrogate data generated by the smoothed version \cite{ByMaGr14,marcottetop} of the Karma model \cite{Karma1993,karma94},
\begin{equation}\label{eq:rde}
	\partial_t{\bf w} = D\nabla^{2}{\bf w} + {\bf f}({\bf w}),
\end{equation}
where ${\bf w}=[u,v]$, $u$ is the (fast) voltage variable, $v$ is the (slow) gating variable,
\begin{align}\label{eq:karmakinetics}
	f_1&=(u^* - v^{M})\{1 - \tanh(u-3)\}u^{2}/2 - u,\\
	f_2&=\epsilon\left\{\beta \Theta_s(u-1) + \Theta_s(v-1)(v-1) - v\right\},\nonumber
\end{align}
and $\Theta_s(u)=[1+\tanh(su)]/2$.
Here $\epsilon$ describes the ratio of the fast and slow time scales, $s$ is the smoothing parameter, and the diagonal matrix $D$ of diffusion coefficients describes the spatial coupling between neighboring cardiac cells (cardiomyocytes). 
The parameters of the model are $M=4$, $\epsilon = 0.01$, $s = 10$, $\beta = 1.389$, $u^{*} = 1.5415$, $D_{11} = 4.0062$, and $D_{22} = 0.20031$, with the length scale corresponding to the size of a cardiomyocyte. 
Along with the Mitchell-Schaeffer model \cite{mitchell2003two}, this is one of the simplest models of excitable media that develops sustained spiral wave chaos from an isolated spiral wave through the amplification of the alternans instability.

\subsection{Analysis}

The method described here uses the voltage variable $u$ (normalized to the range [0,1] for convenience) to reconstruct the position of PSs.
In some instances, e.g., electrophysiology studies using basket catheters, the voltage data are only available on a coarse spatial grid.
To enable rotor mapping with meaningful accuracy in such cases, the data at each frame of the recording are mapped onto a sufficiently fine mesh using bicubic interpolation.

Following the original study that introduced the topological approach \cite{marcottetop}, we will define PSs as intersections of two smooth curves $\ell_1$ and $\ell_2$. For data corresponding to transmembrane voltage (obtained from a model or optical mapping using a voltage-sensitive dye), these curves can be conveniently defined as the zero level sets of $\dot{u}$ and $\ddot{u}$, which form the boundaries $\partial R$ and $\partial E$, respectively, of the refractory region
\begin{align}
R=\{(x,y) : \dot{u}<0\}
\end{align}
and the excited region
\begin{align}
E=\{(x,y) : \ddot{u}<0\}.
\end{align}
Given that our data are discrete and noisy, properly defining the level sets and PSs requires some care.
$\partial R$ and $\partial E$ correspond to peaks and troughs of $u$ and $\dot{u}$, respectively, where the temporal derivative can be computed using finite differencing of the discrete signal.
In the presence of noise (i.e., when dealing with experimental recordings), the data may be smoothed using a Gaussian kernel with spatial and temporal widths $\sigma_s$ and $\sigma_t$, respectively, before the time derivative is computed. Even after smoothing, $u$ and especially $\dot{u}$ can remain noisy. So each peak or trough is required to have a minimum prominence (a fraction $MPP_1$ or $MPP_2$ of the range of $u$ or $\dot{u}$, respectively, selected based on the overall level of noise, as described in the Appendix) and separation (generally a fixed fraction $\delta$ of the dominant period of oscillation) from the nearest peak/trough.

\begin{figure}[t]
\subfigure[]{\frame{\includegraphics[width=0.47\columnwidth]{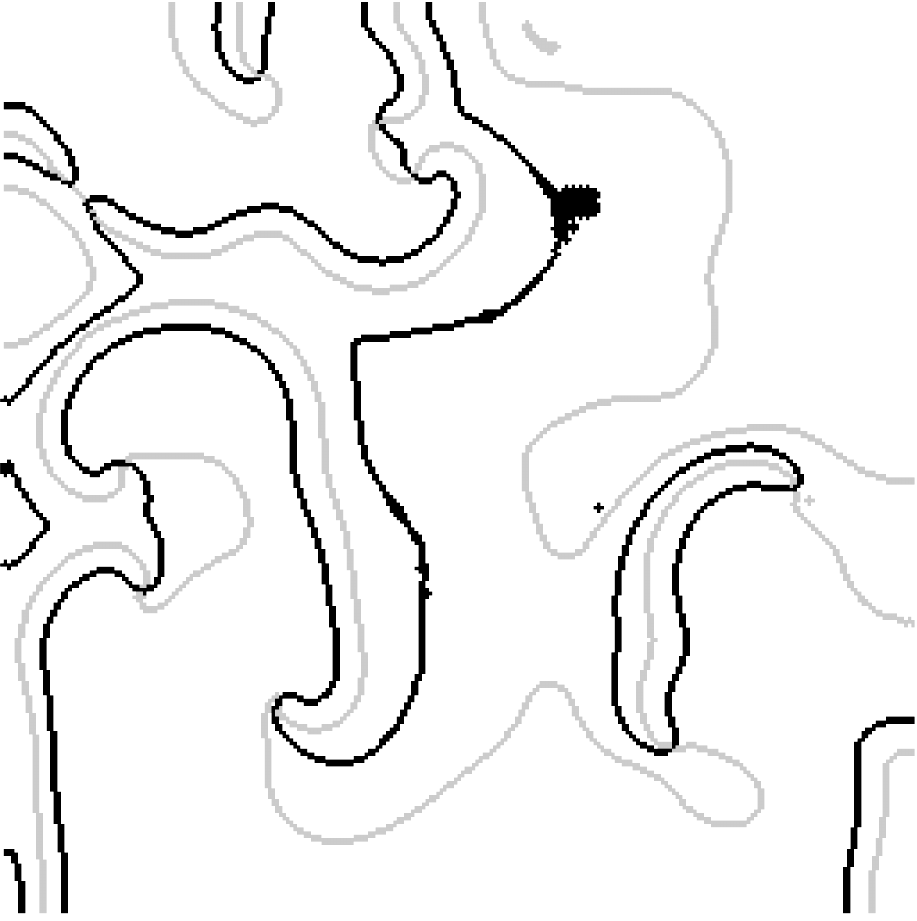}}} \hspace{2mm}
\subfigure[]{\frame{\includegraphics[width=0.47\columnwidth]{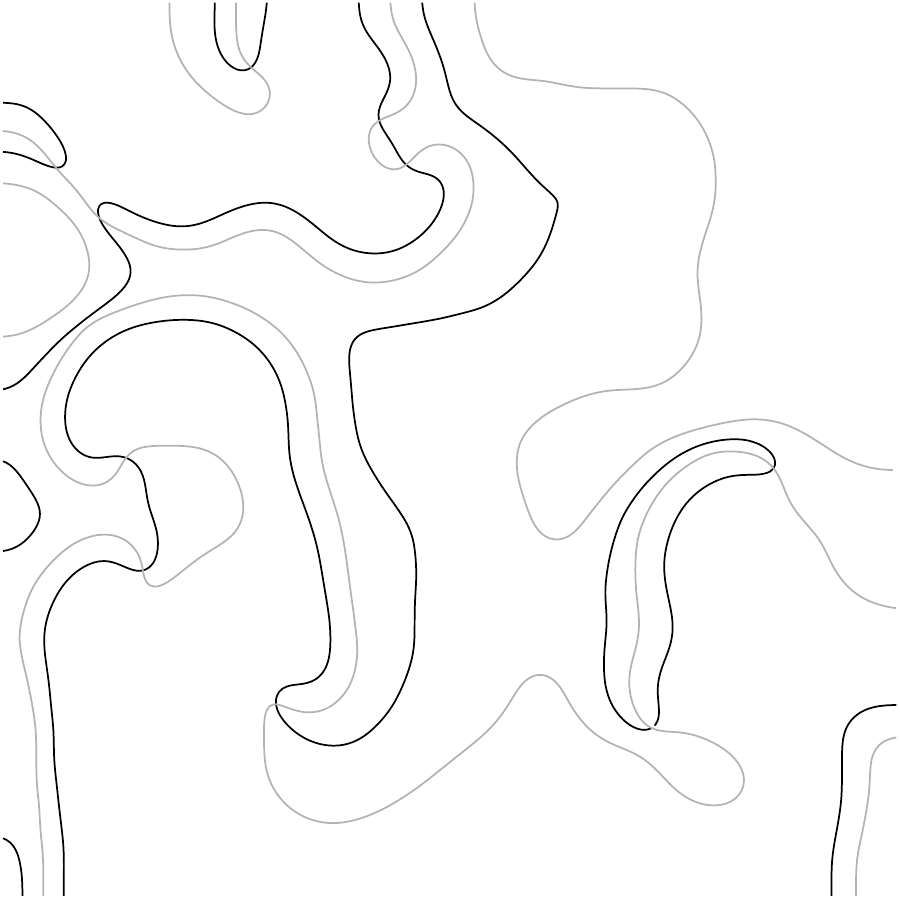}}}
\caption{Comparison of sets $\ell^k$ (a) and $\bar{\ell}^k$ (b) for noiseless data, with $k=1$ in black and $k=2$ in gray. The domain is 256 $\times$ 256 grid points. Panel (a) contains a number of artifacts, which have to do with a characteristic feature of most cardiac models, namely very flat repolarization plateaus.
}
\label{fig:lvl}
\end{figure}

Spatial discreteness of the data -- we assume $u$ is measured on a uniform grid $(x_i,y_j,t_n)$ in space and time -- limits the accuracy with which the level sets $\partial R$ and $\partial E$ can be defined based on local data to the resolution of the underlying spatial grid. For instance, if we simply identify the grid points that correspond to the peaks or troughs in each frame, we end up with a sparse set of isolated points. In order to define a continuous set that can be considered a discrete approximation of $\partial R$ and $\partial E$, we will use the following procedure. Let $t^1_{ijm}$ denote the position of subsequent peaks (troughs) of $u(x_i,y_j,t)$ for $m$ even (odd), and define $t^2_{ijm}$ analogously for $\dot{u}(x_i,y_j,t)$.
Further, let us define a noise-insensitive analog of the sign of the time derivative
\begin{align}\label{eq:s1}
s^k_{ijn} =  \begin{cases}
       1,&  t^k_{ijm-1}<t_n\leq t^k_{ijm}\\
       -1,& t^k_{ijm}<t_n\leq t^k_{ijm+1}
    \end{cases}
\end{align}
for some even number $m$ and $k=1$ or 2.
Then the sets
\begin{align}\label{eq:ls2}
\ell^k(t_n)=\{(x_i,y_j) : \exists i', j' :\ &s^k_{ijn}s^k_{i'j'n}<0,\nonumber \\ 
&|i-i'|+|j-j'|=1\}
\end{align}
are discrete generalizations of the level sets $\partial R$ and $\partial E$ with a minimal width of 2 grid points and no gaps, as illustrated in Fig. \ref{fig:lvl}(a).

\begin{figure}[t]
\centerline{\includegraphics[width=\columnwidth]{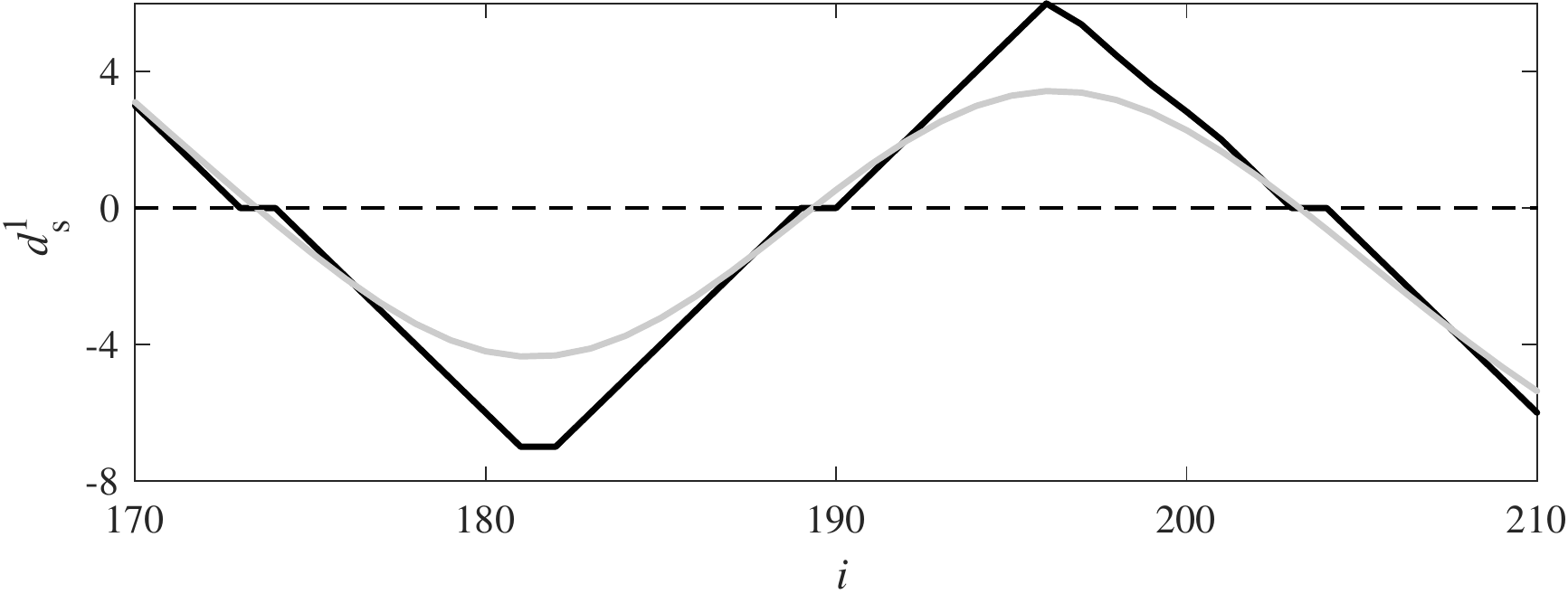}}
\caption{The signed distance $d^1_s$ (black) and its smoothed version $\bar{d}^1_s$ (gray) at a fixed time over a $j=\mathrm{const}$ slice of the domain.
}
\label{fig:df}
\end{figure}

In order to define the position of the level sets with sub-grid precision using global, rather than local (and hence noisy), information we use the following approach.
For each of $\ell^1$ and $\ell^2$, unsigned distance functions $d^1$ and $d^2$, respectively, are constructed using the MATLAB function {\tt bwdist} \cite{maurer2003linear}. 
These are in turn converted into signed distance functions
\begin{align}
d_s^k(x_i,y_j,t_n) = -s^k_{ijn}d^k(x_i,y_j,t_n),
\end{align}
which incorporate global information across the entire domain.
Next, in order to ultimately smooth and sharpen the level sets, a spatial convolution of the signed distance functions with a Gaussian kernel with spatial width $\sigma_d$ is computed, yielding a pair of smoothed distance fields $\bar{d}^k_s$. 
A comparison of the smoothed and unsmoothed versions of the signed distance function is presented in Fig. \ref{fig:df}.
Note the plateaus in the unsmoothed distance function, representing the finite thickness of the discrete set of grid points from which it is computed.

Now we can finally define curves $\bar{\ell}^k$ as the zero level sets of the smoothed distance fields $\bar{d}^k_s$ using the MATLAB function {\tt contour}.
These curves are piecewise continuous and smooth, although in practice they are parametrized by a sequence of connected points in $\mathbb{R}^2$.
Note that $\ell_k$ define the true positions of $\partial R$ and $\partial E$ with a precision of one grid point or better. 
As Fig. \ref{fig:lvl} illustrates, for noiseless data, $\bar{\ell}^k$ provide accurate representations of $\ell^k$ and hence $\partial R$ and $\partial E$, despite the relatively aggressive smoothing.

In practice, we will determine PSs as the intersections of $\bar{\ell}^1$ and $\bar{\ell}^2$, which are computed with sub-grid precision using the MATLAB function {\tt intersections} \cite{intersections}. 
The chirality (or topological charge) $q$ of each PS can be computed using the gradients of $\bar{d}^1_s$ and $\bar{d}^2_s$, which are nearly constant in the vicinity of a PS, as follows:
\begin{align}
q = \mathrm{sign}(\hat{\mathbf z} \cdot \nabla \bar{d}^1_s \times \nabla \bar{d}^2_s),
\end{align}
These gradients are approximated at the four grid points nearest to the PS using finite differencing and then interpolated to the exact location of the PS.
This interpolation can be essential to correctly determine the chiralities of a pair of phase singularities separated by only a few grid points (in practice, as few as one).

Chirality plays an important role in the topological analysis of the excitation patterns produced during arrhythmias \cite{marcottetop}. 
It is also useful for reconstructing PS trajectories, which are computed using a MATLAB implementation \cite{track} of the IDL particle tracking method \cite{crocker1996}, with the positions and chiralities of all PSs found at each time step as input parameters. 
Both experimental and numerical data feature many short trajectories that correspond to virtual spiral waves that exist for a fraction of a rotation period. 
Such structures do not appear to play a dynamically important role, so in our analysis we ignore PSs with lifetimes shorter than the dominant period of oscillation.
For comparison, during electrophysiological studies in a clinical setting, only spiral waves that persist for at least two rotations are considered \cite{ashikaga2018}.

\section{Results}
\label{sec:results}

To test the algorithm, we generated 4000 frames of surrogate data (after discarding 600 frames representing the initial transient) separated by one time unit by numerically integrating the model described above on a square domain of size $256 \times 256$ with no-flux boundary conditions.
This can be considered a fine mesh as it fully resolves all of the spatial features of the solution.
In our units, the typical rotation period of a spiral wave is $T=53$, the mean separation between PSs is $L=46$, and the mean number of PSs is 12.7.

\subsection{Benchmark}

\begin{figure}[t]
\subfigure[]{\includegraphics[width=0.47\columnwidth]{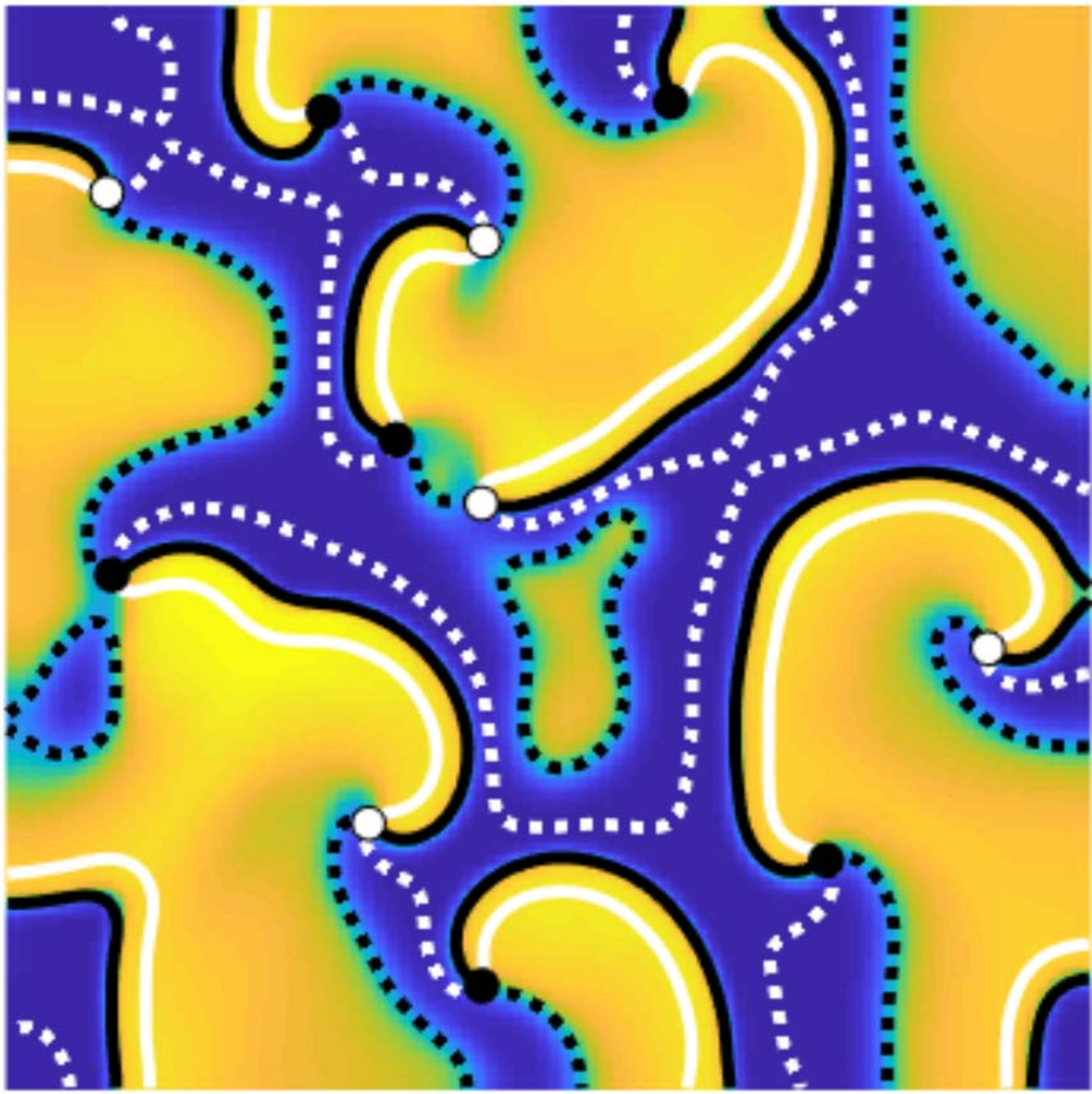}} \hspace{1mm}
\subfigure[]{\includegraphics[width=0.47\columnwidth]{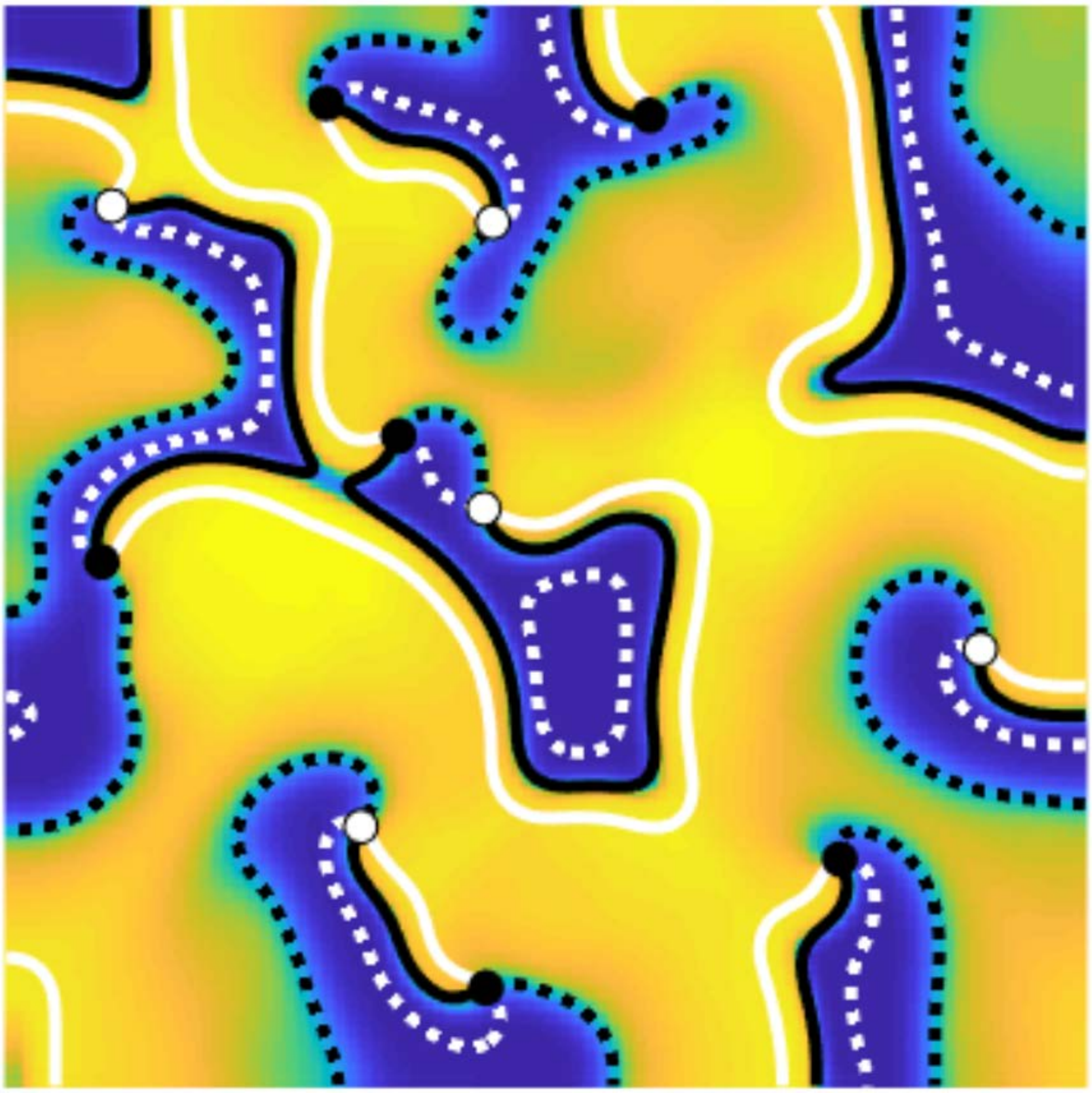}} \\
\subfigure[]{\includegraphics[width=0.47\columnwidth]{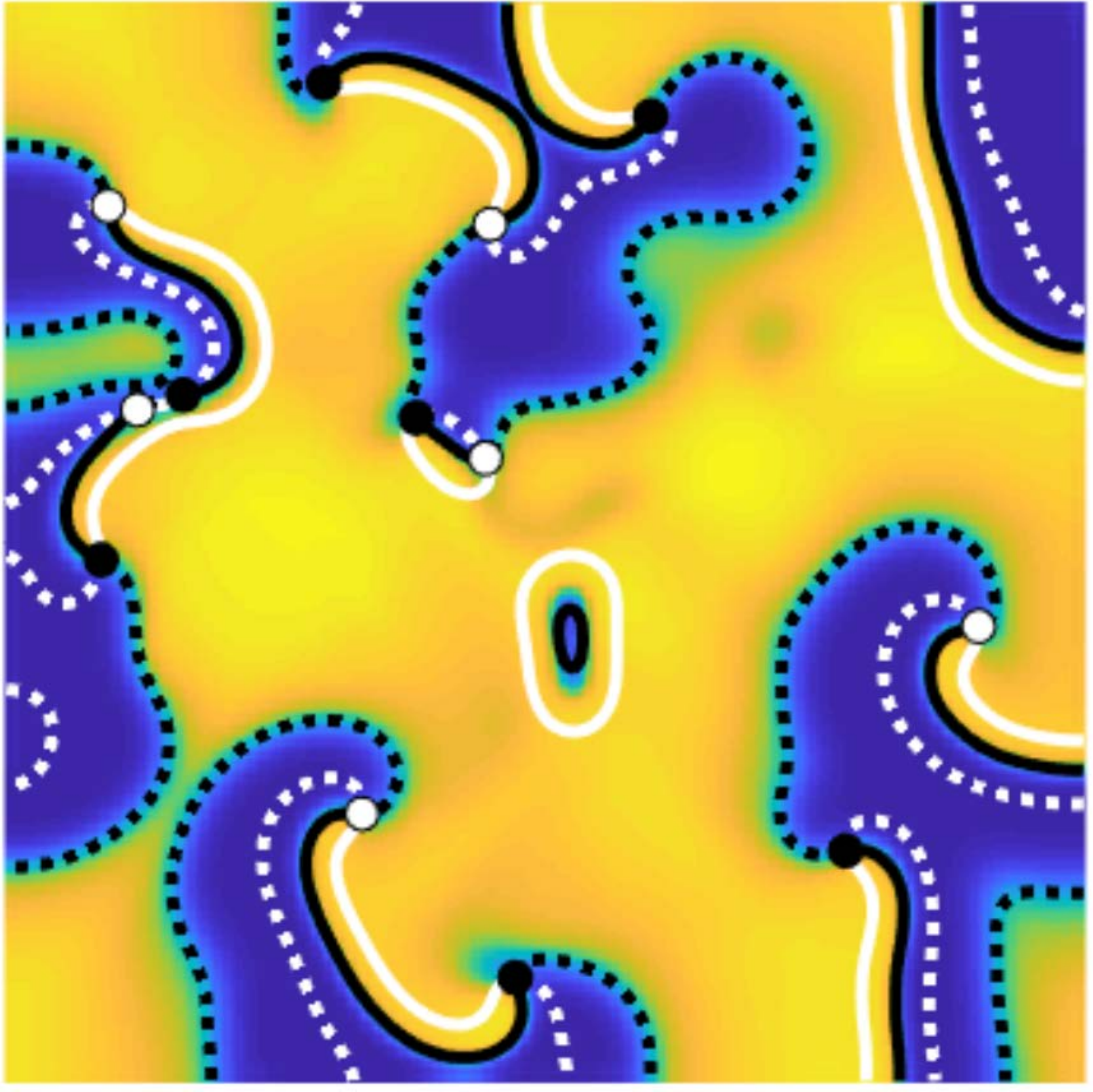}} \hspace{1mm}
\subfigure[]{\includegraphics[width=0.47\columnwidth]{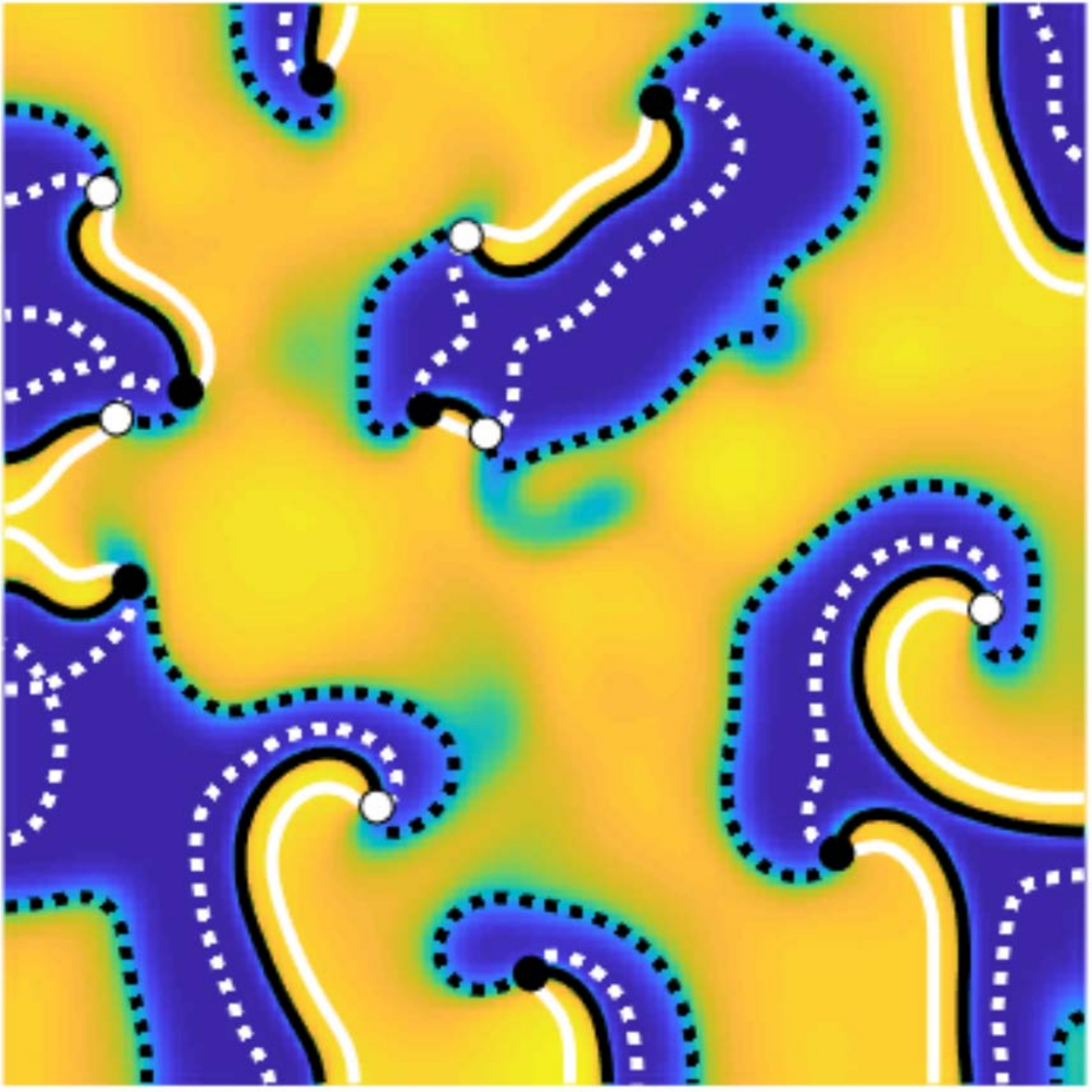}} \\
\subfigure[]{\includegraphics[width=0.47\columnwidth]{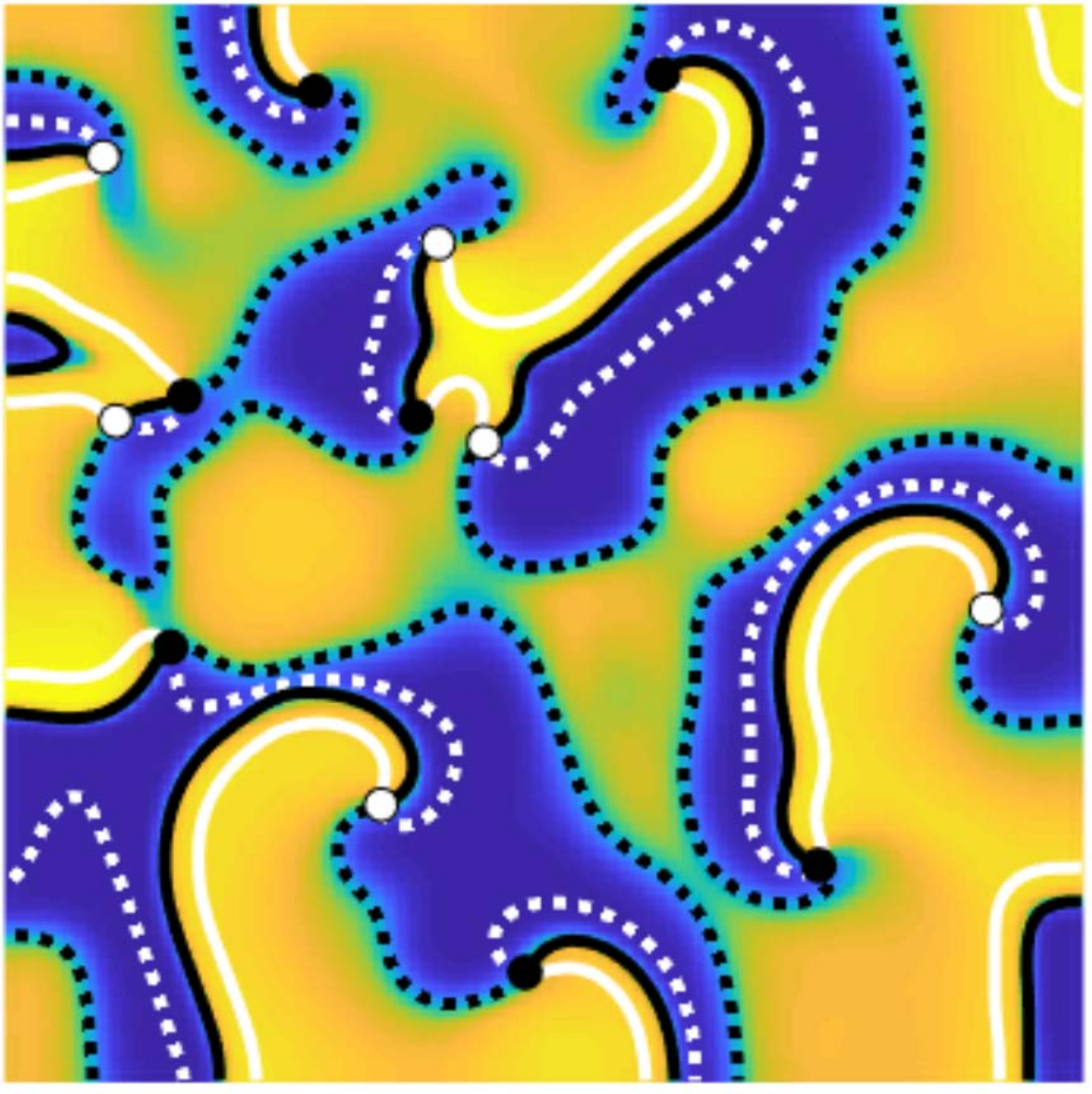}} \hspace{1mm}
\subfigure[]{\includegraphics[width=0.47\columnwidth]{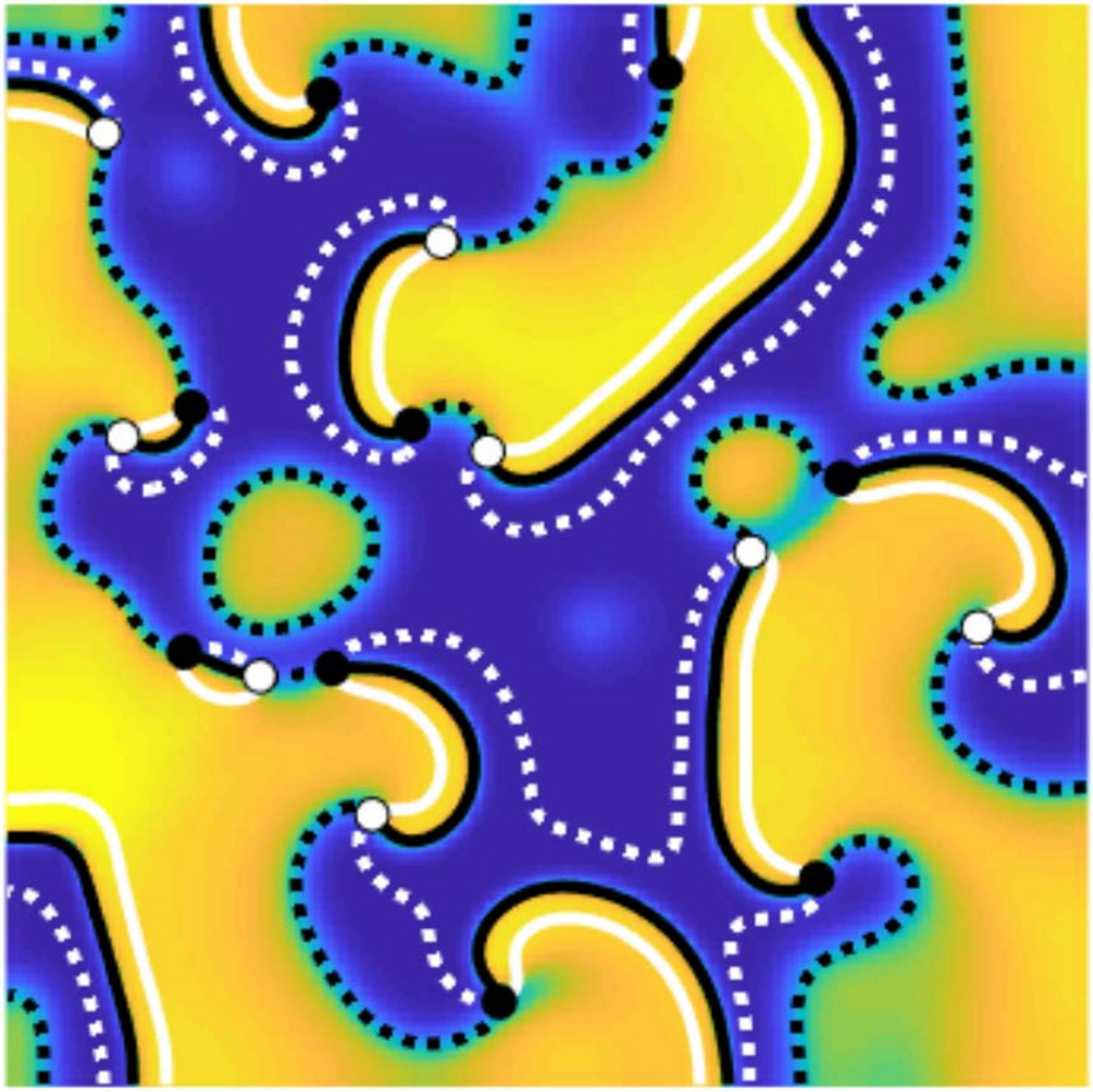}}
\caption{Snapshots of benchmark data (colorbar is shown in Fig. \ref{fig:g}(a)) equally spaced in time over 56 frames (approximately one rotation period), with curves $\bar{\ell}^1$ (white) and $\bar{\ell}^2$ (black) and PSs superimposed. Here and below, solid and dashed white segments correspond to the leading and trailing edges of the refractory region, respectively; solid and dashed black correspond to the wavefront and waveback. PSs with chirality +1 and -1 are respectively shown as black and white circles. The $x$ ($y$) axis is horizontal (vertical). 
A full movie is provided in the supplementary material.}
\label{fig:f}
\end{figure}

\begin{figure}[t]
\centerline{\includegraphics[width=\columnwidth]{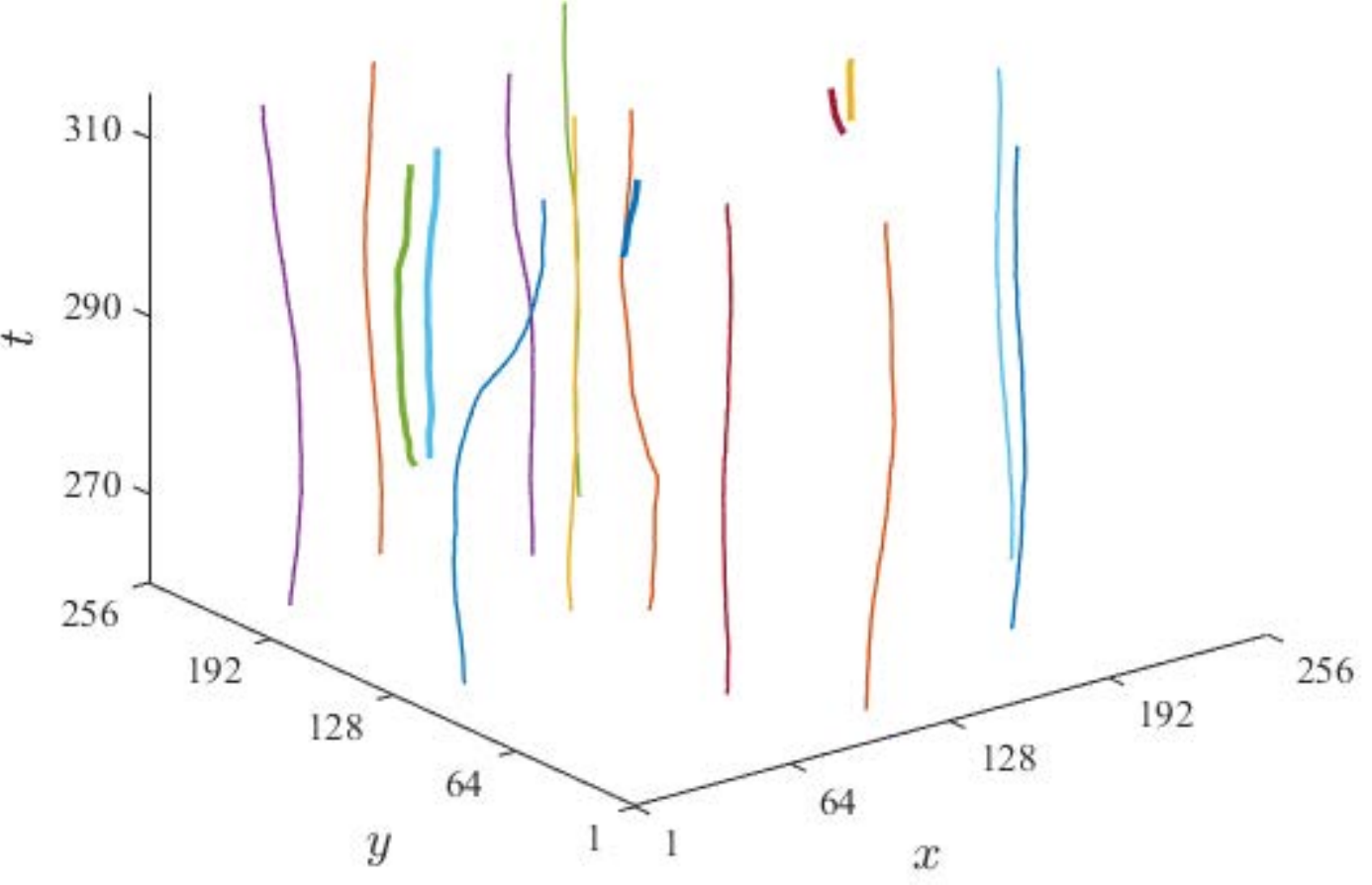}}
\caption{Trajectories of PSs with lifetimes of at least one period during the same time interval as shown in Fig. \ref{fig:f}. Thicker curves correspond to PSs created during this period.
}
\label{fig:tracks}
\end{figure}

In order to establish a benchmark for quantifying how our approach copes with noise and data sparsity, we analyzed the data using a parameter set representing minimal smoothing (see Appendix). 
Fig. \ref{fig:f} presents six equally spaced snapshots of the benchmark (voltage) data over roughly a rotation period. 
Superimposed are the curves $\bar{\ell}^1$ (the boundary of the refractory region, in white) and $\bar{\ell}^2$ (the boundary of the excited region, in black). 
The intersections define PSs (black/white circles for positive/negative chirality).

Many spiral waves are seen to rotate stably around fixed or weakly meandering PSs.
The trajectories of the long-lived PSs are shown over the same interval in Fig. \ref{fig:tracks}, with the thicker curves corresponding to PSs created during this time.
Three different PS pair creation events are visible during this period: one between snapshots (b) and (c) near the left edge of the domain, and two between snapshots (e) and (f) near the left and right edges. 

Notably, the lone thick blue trajectory in Fig. \ref{fig:tracks} appears to be missing its opposite chirality counterpart.
In fact, that counterpart is not shown since it corresponds to a short-lived ``virtual" PS which annihilates with a nearby long-lived PS soon after the last frame in Fig. \ref{fig:f}.
Note that the trajectories of created PS pairs do not start at the exact same point because of the finite temporal resolution of the data.
When the PSs are created and destroyed, they move especially quickly, separating by several grid spacings in one time unit.
Much higher temporal resolution is needed to resolve the fast motion of PSs during pair creation/annihilation events.

\begin{figure}[t]
\subfigure[]{\includegraphics[width=0.47\columnwidth]{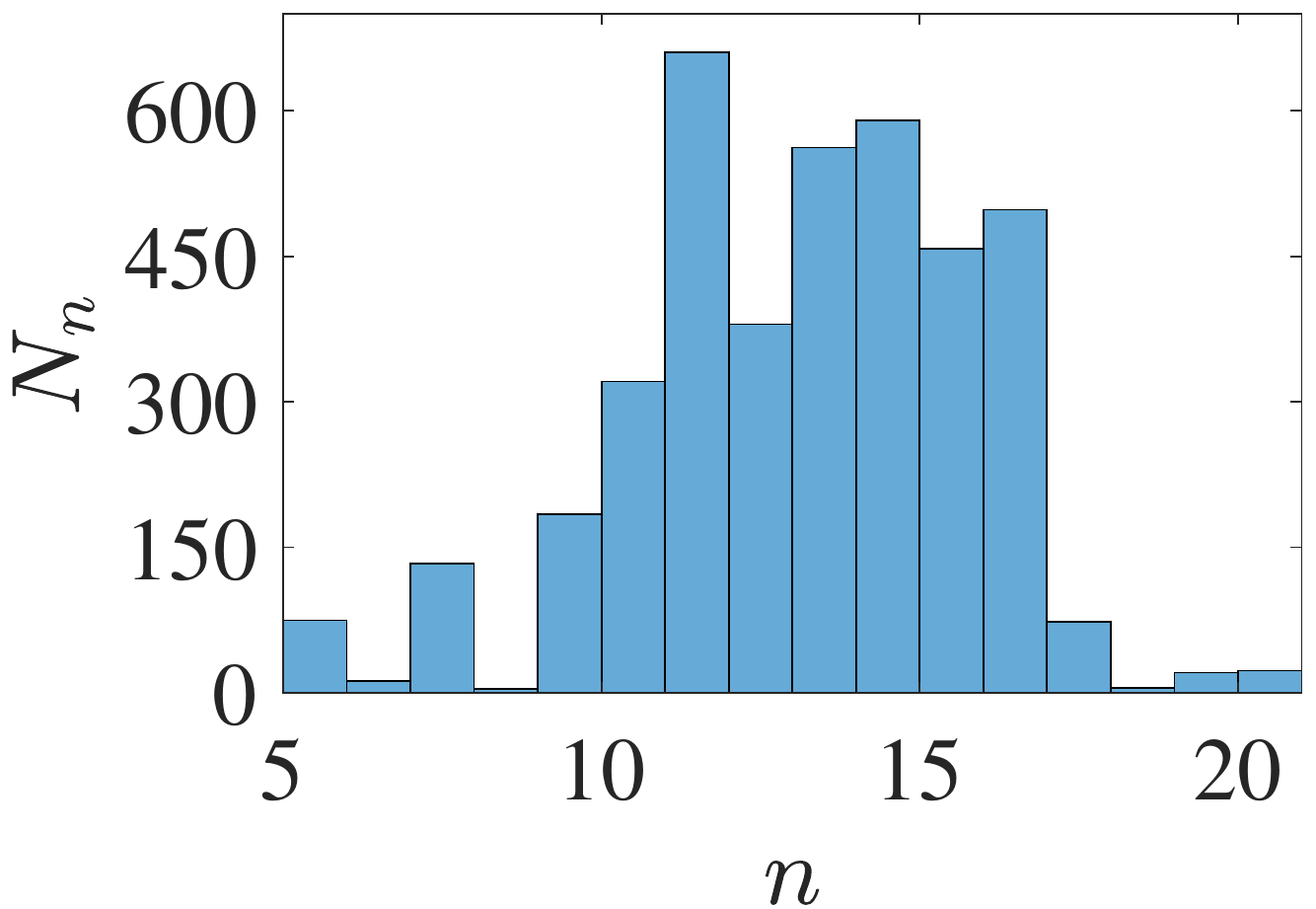}} \hspace{3mm}
\subfigure[]{\includegraphics[width=0.45\columnwidth]{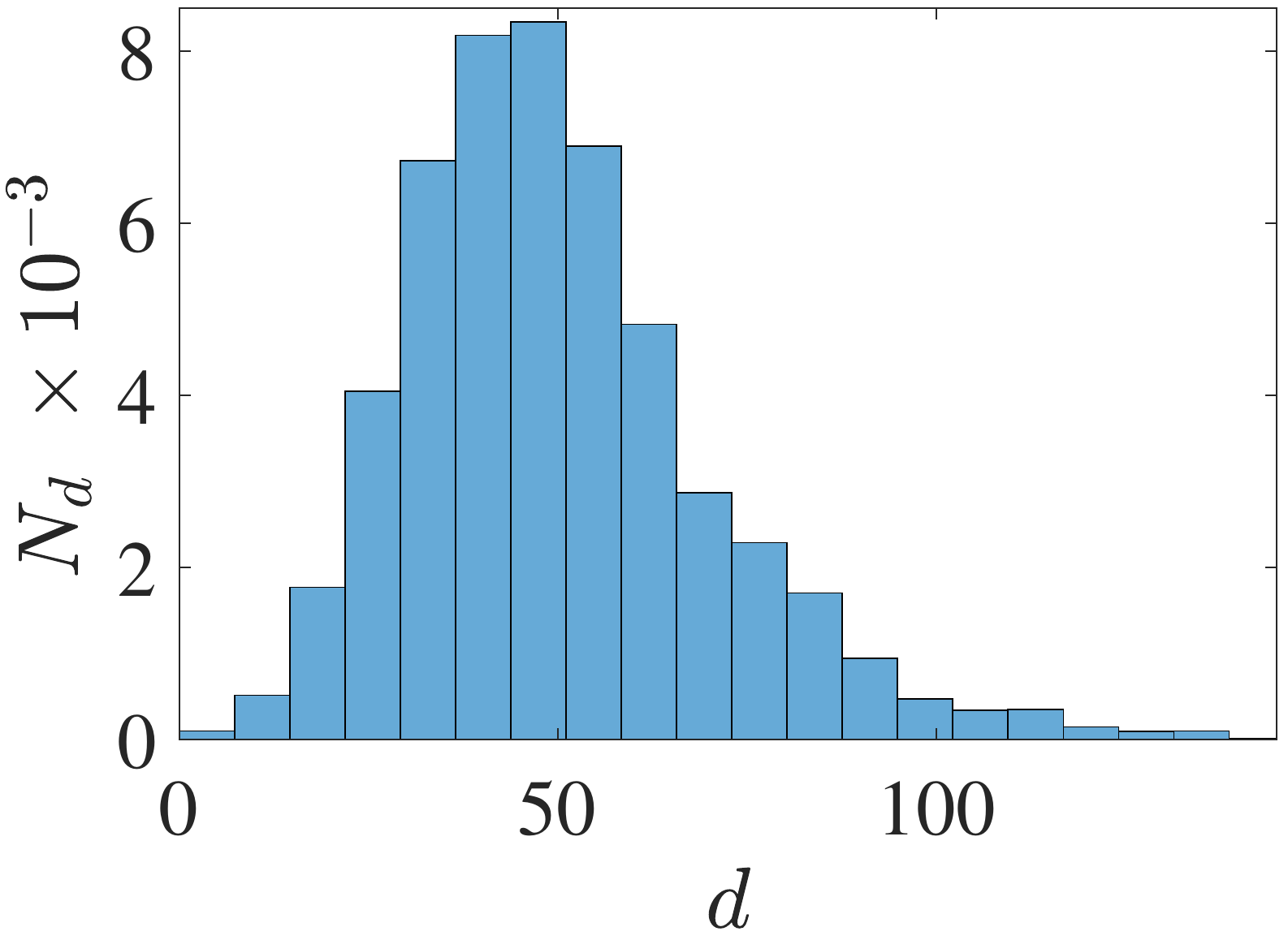}} \\
\subfigure[]{\includegraphics[width=0.47\columnwidth]{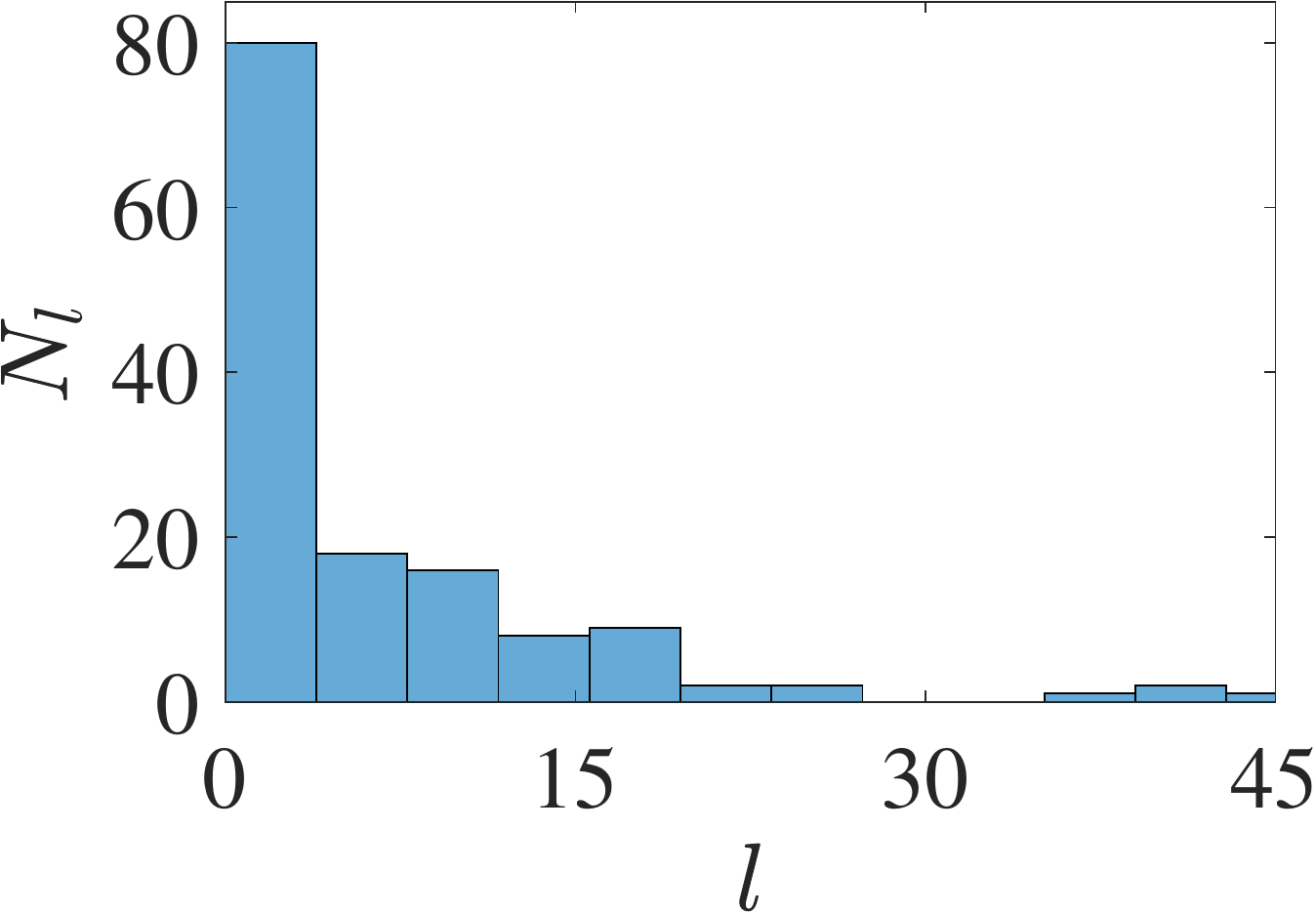}} \hspace{1mm}
\subfigure[]{\includegraphics[width=0.47\columnwidth]{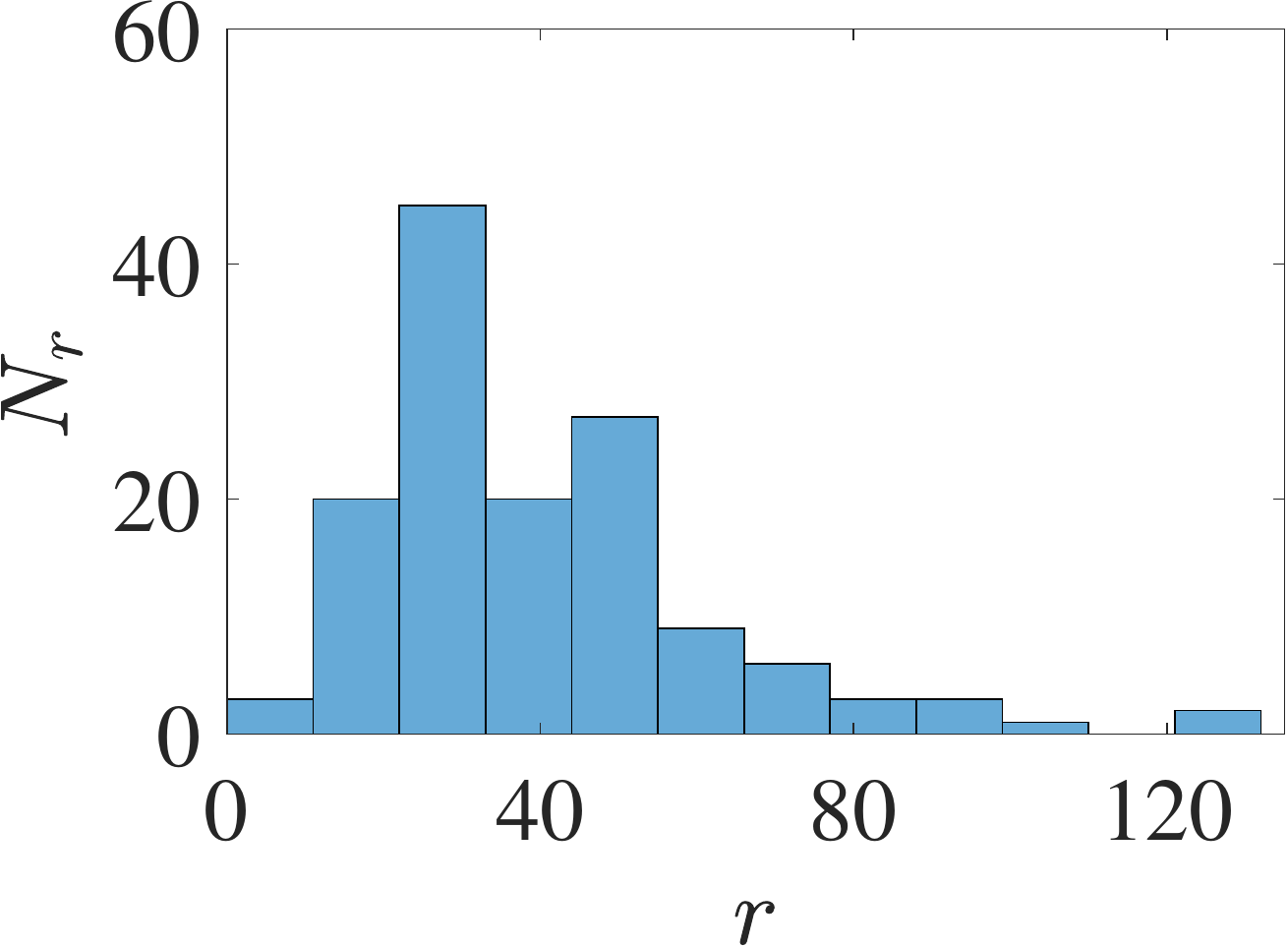}}
\caption{PS statistics. (a) Histogram of the total number $n$ of PSs in each frame. (b) Histogram of the distance $d$ from each PS to the nearest PS of opposite chirality, computed separately for each PS in each frame. (c) Histogram of the lifetime $l$ in periods of each PS. (d) Histogram of the separation $r$ between the most distant pair of points along the trajectory of each PS.
}
\label{fig:hist}
\end{figure}

The ability of our algorithm to automatically track PSs (both short- and long-lived) allows one to generate statistics that could be extremely useful (e.g., for model validation) but would be hard to obtain otherwise.
To illustrate this, Fig. \ref{fig:hist} shows various PS statistics (only taking spiral waves that complete at least one revolution into account) over the course of the entire simulation.
In particular, we find that the number of PSs ranges rather widely (between 5 and 20, as shown in panel (a)). 
This illustrates that our approach can easily and reliably identify at least 20 PSs simultaneously.
The distances between PSs also vary rather significantly, with a pronounced peak at 46 units (panel (b)).
As we will show below, it is this characteristic length scale, not the wavelength of the pattern (on the order of $90$ units here), that determines the sparsity at which our method starts to break down.

We find that while the majority of spiral waves are relatively short lived (panel (c)), some can live for up to 45 periods (for reference, the duration of the entire data set corresponds to 75 periods).
Such instances of functional reentry could easily be mistaken for structural reentry in a clinical setting.
In light of this ``longevity,'' it is perhaps not surprising that some PSs drift over distances exceeding half the size of our rather large system (panel (d)).
While the lifetime and drift statistics of PSs in the Karma model may not be particularly relevant for atrial fibrillation, a similar analysis of data from basket catheters could yield a treasure trove of clinically valuable information.

\subsection{Sensitivity to noise and sparsification}

To represent the effect of imperfections in realistic experimental recordings, noisy data sets were produced by adding random Gaussian-distributed white noise with some standard deviation $\eta$ to the benchmark data.
In addition, sparsified data were generated by sampling from the benchmark or noisy data on a uniform grid with lower resolution by a factor of $2^n$ in each spatial dimension for some integer $n>1$.
The sparsified data were then interpolated back onto the original grid for processing.

\begin{figure}[t]
\includegraphics[width=0.99\columnwidth]{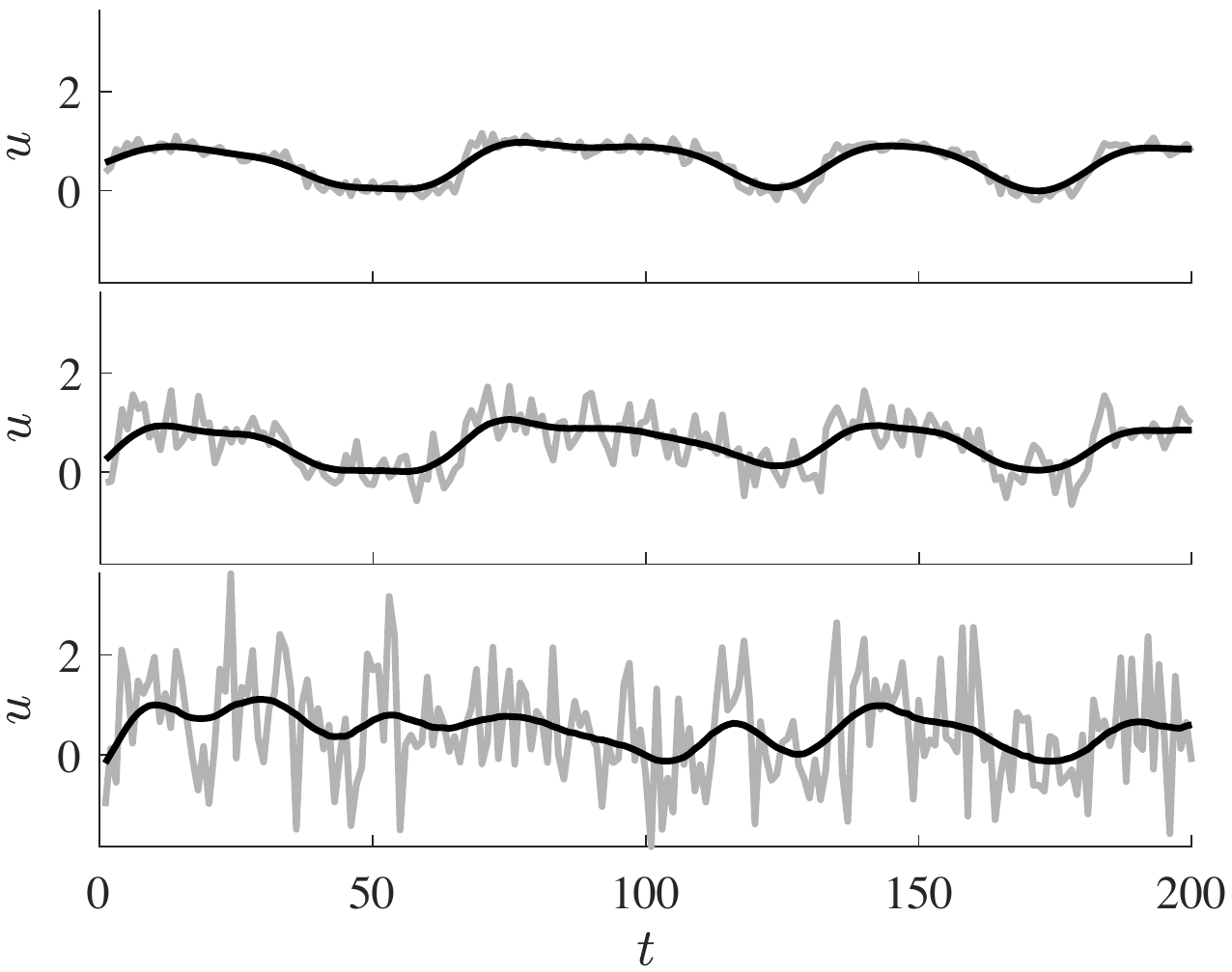}
\caption{A typical time trace of the voltage signal before (gray) and after (black) temporal smoothing for different noise levels (from top to bottom, $\eta=0.1$, $\eta=0.3$, and $\eta=1$).}
\label{fig:u}
\end{figure}

For all noise and sparsity levels, the resulting data were processed using modified parameter sets mildly optimized to deal with high levels of noise (cf. Appendix). 
Note that, in order to preserve precision at different levels of sparsification, no initial spatial smoothing is applied (the raw and temporally smoothed signals are compared in Fig. \ref{fig:u}).
The robustness of the algorithm is thus ensured without relying on averaging of high-spatial-resolution data to counteract the effects of noise.
As we show below, our results are robust to both noise and sparsification but show some minor reduction in precision of PS location, even for the benchmark data.

\begin{figure}[t]
\subfigure[]{\includegraphics[width=0.47\columnwidth]{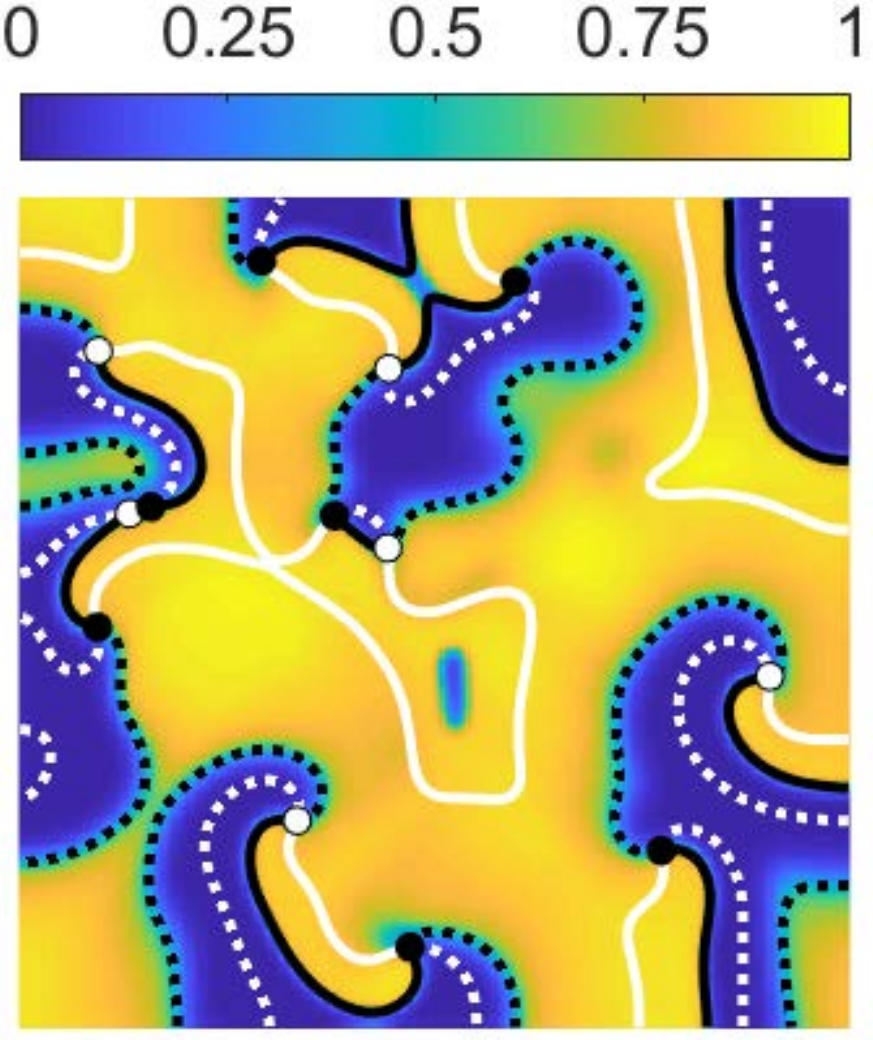}} \hspace{1mm}
\subfigure[]{\includegraphics[width=0.47\columnwidth]{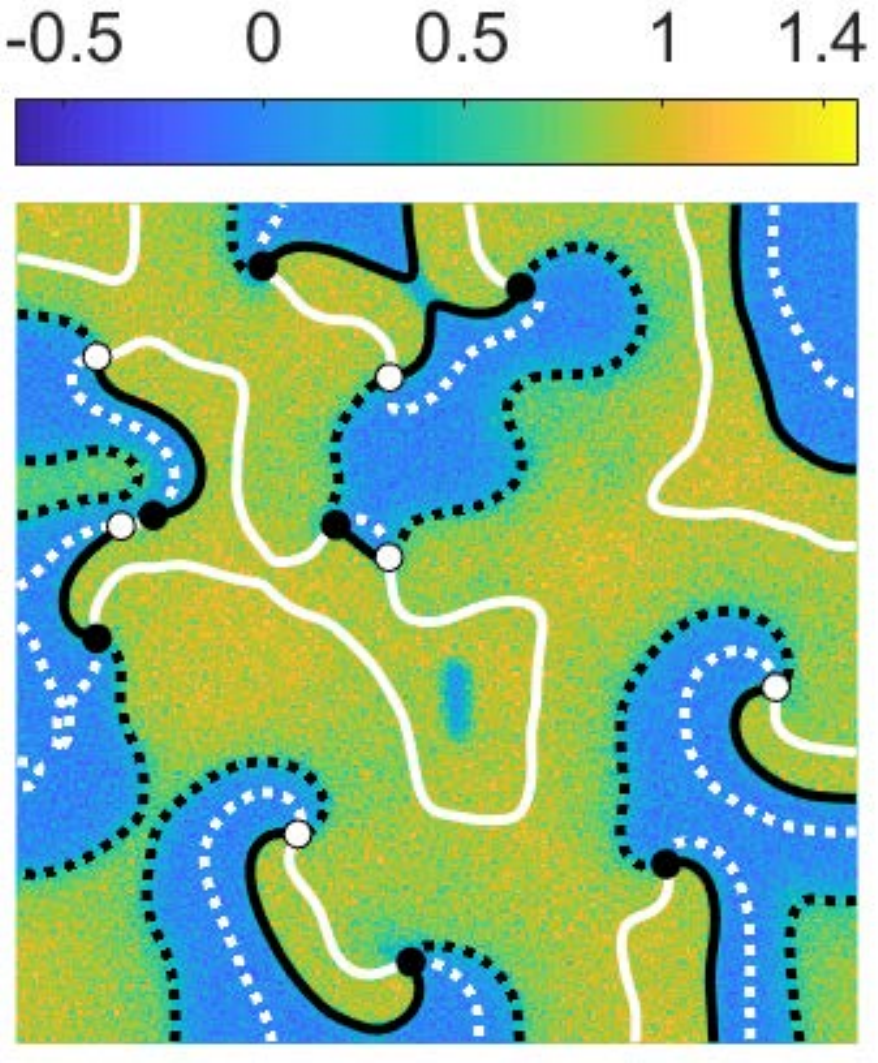}} \\
\subfigure[]{\includegraphics[width=0.47\columnwidth]{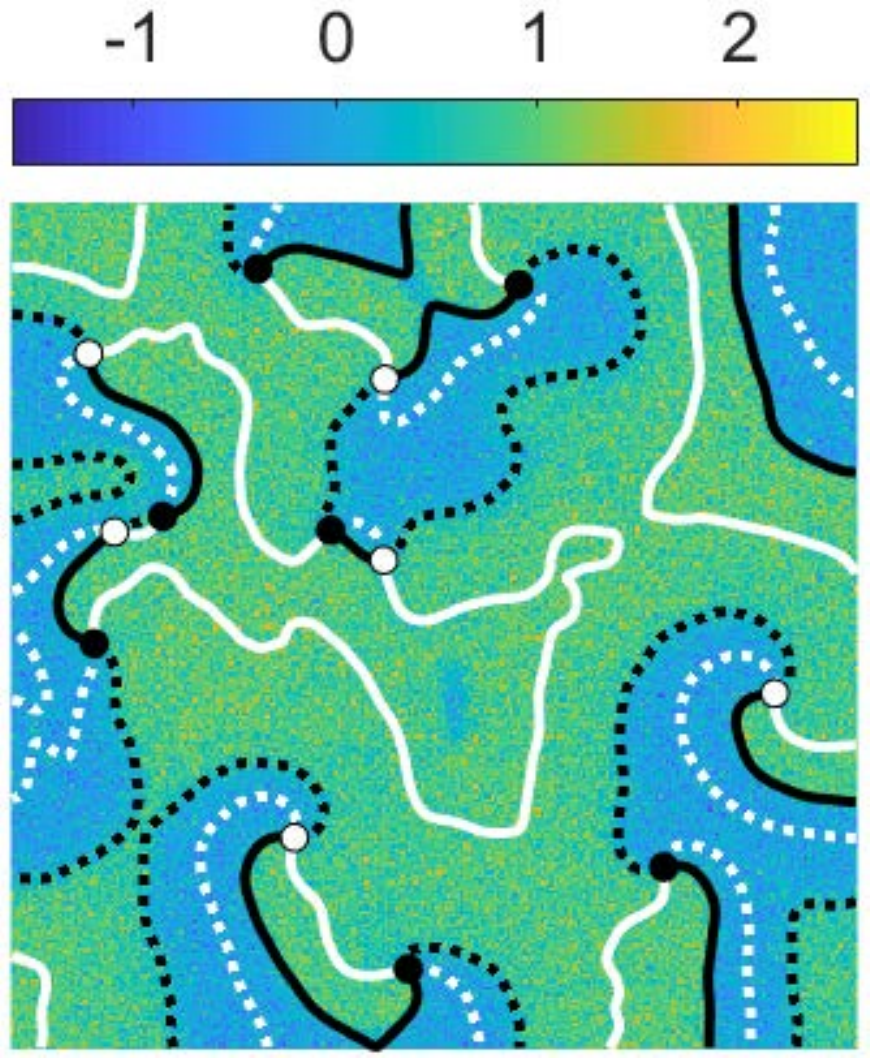}} \hspace{1mm}
\subfigure[]{\includegraphics[width=0.47\columnwidth]{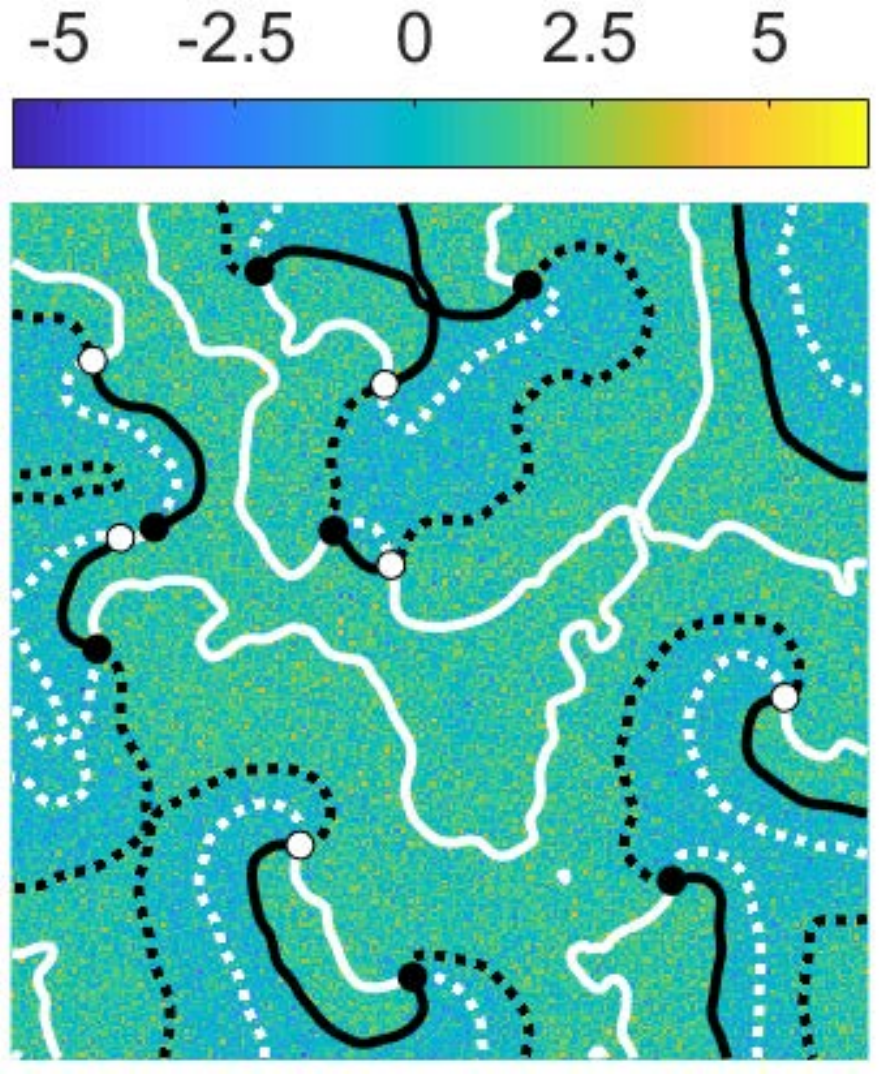}}
\caption{The frame shown in Fig. \ref{fig:f}(c) with four different levels of added noise ($\eta=0,0.1,0.3,1$); overlaid are the curves $\bar{\ell}^1$ and $\bar{\ell}^2$ and PSs computed from the noisy data in each case.
}
\label{fig:g}
\end{figure}

The effect of noise alone on the performance of the algorithm is illustrated in Fig. \ref{fig:g}, which shows the level sets $\bar{\ell}^1$, $\bar{\ell}^2$, and the PSs computed for the same frame with four levels of Gaussian white noise (standard deviation $\eta=0,0.1,0.3,1$).
As the data quality deteriorates, the computed curves, especially $\bar{\ell}^1$, become increasingly unreliable; it is impossible to filter out all false peaks in the data without ignoring the smaller but legitimate and dynamically important fluctuations in voltage.
Nevertheless, their intersections (the PSs) are located with high precision in all cases, even when $\eta$ is as large as the entire range of the original data.

\begin{figure}[t]
\subfigure[]{\includegraphics[width=0.47\columnwidth]{g1.pdf}} \hspace{1mm}
\subfigure[]{\includegraphics[width=0.47\columnwidth]{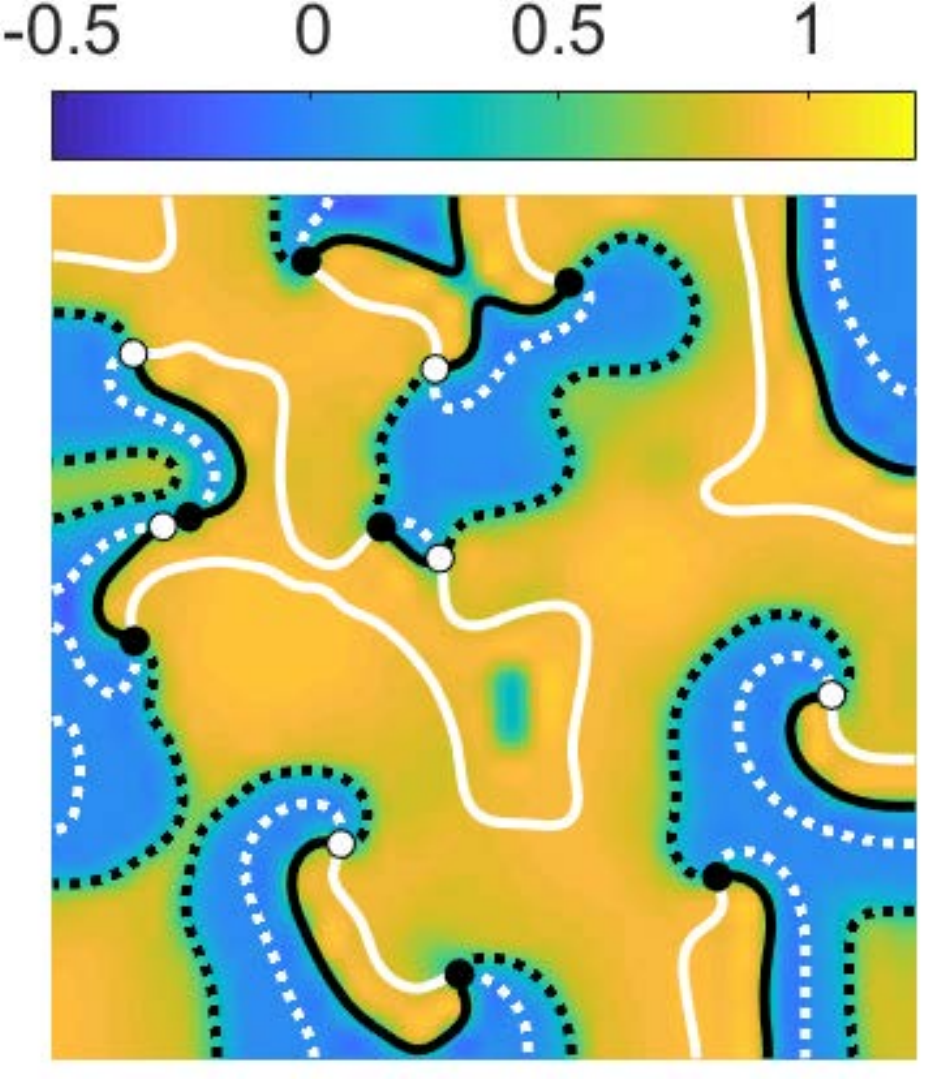}} \\
\subfigure[]{\includegraphics[width=0.47\columnwidth]{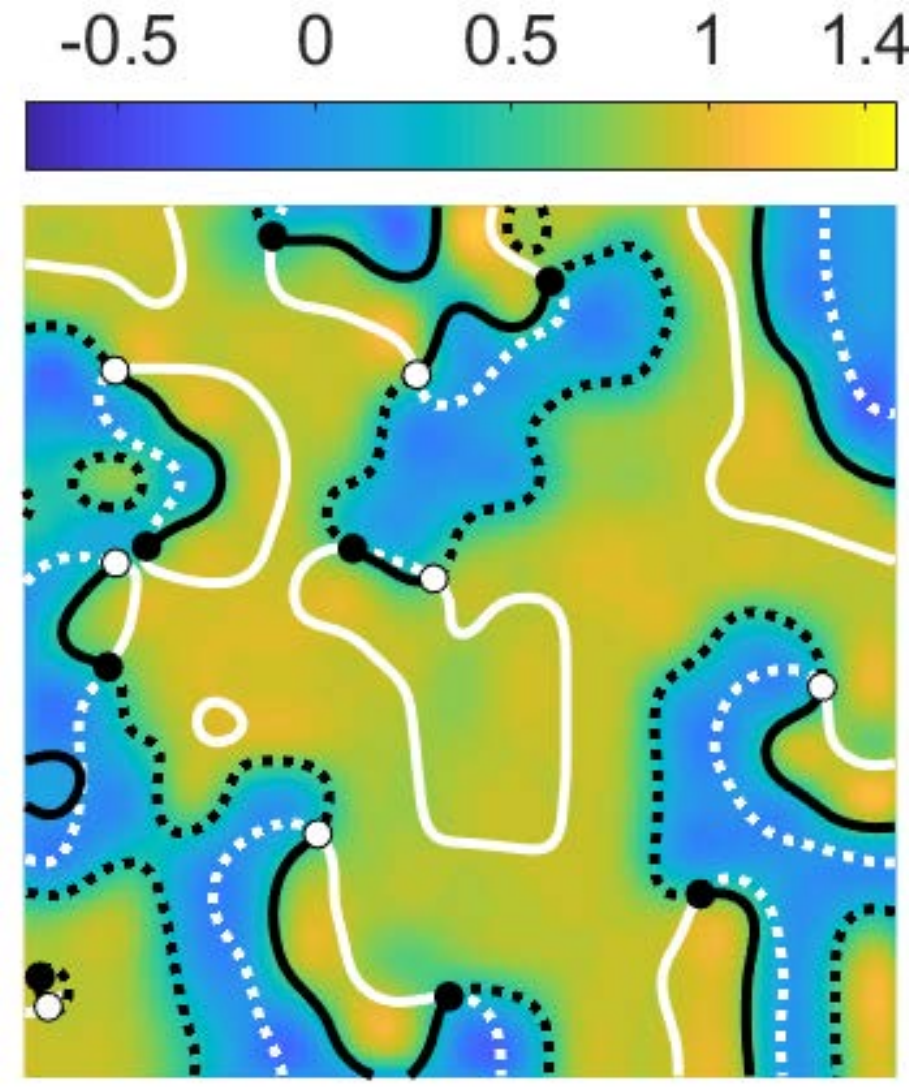}} \hspace{1mm}
\subfigure[]{\includegraphics[width=0.47\columnwidth]{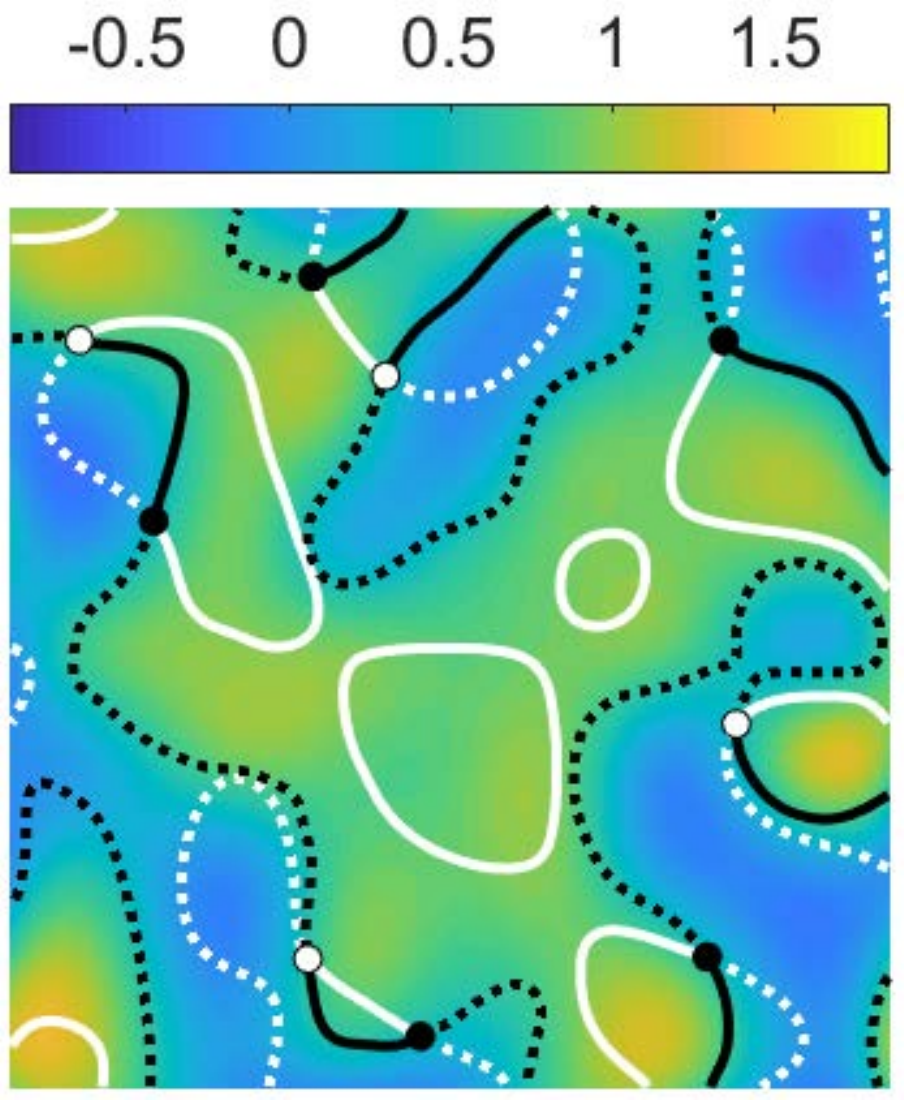}}
\caption{The frame shown in Fig. \ref{fig:f}(c) after interpolating from four levels of sparsification: (a) spatial resolution 256 $\times$ 256 (same as benchmark), (b) 32 $\times$ 32, (c) 16 $\times$ 16, (d) 8 $\times$ 8. As in the previous figure, the curves $\bar{\ell}^1$ and $\bar{\ell}^2$ and PSs computed from the sparsified data are overlaid.
}
\label{fig:h}
\end{figure}

Similarly, the effect of sparsity alone is illustrated in Fig. \ref{fig:h}, which shows the same frame with (noiseless) data interpolated from coarse grids with four different spatial resolutions: (a) $256 \times 256$, (b) $32 \times 32$, (c) $16 \times 16$, (d) $8 \times 8$. 
For resolutions down to $16 \times 16$, the computed level sets are qualitatively very similar to the benchmark and all long-lived PSs are correctly detected and located with precision substantially better than the coarse grid spacing.
In the $16 \times 16$ snapshot, there is a pair of virtual PSs in the lower left corner that are not present in the benchmark analysis. 
These are short-lived and so are discarded in our analysis.
Even for data on an $8\times8$ spatial grid, a large fraction of PSs were correctly identified, with one false positive;
the error in the position of the correct PSs is substantially smaller than $256/8=32$ units.
However, it is apparent that many of the dynamical features cannot be properly resolved when the grid resolution becomes comparable to the mean separation between PSs.

\begin{table}[b]
\begin{tabular}{*{6}{|c}|}
\hline & $256 \times 256$ & $64 \times 64$ & $32 \times 32$ & $16 \times 16$ & $8 \times 8$\\ \hline
$\eta=0$ & 0.995 & 0.995 & 0.994 & 0.955 & \;0.255\;\\ \hline
$\eta=0.1$ & 0.993 & 0.994 & 0.992 & 0.957 & 0.308 \\ \hline
\;$\eta=0.3$\; & 0.988 & 0.988 & 0.985 & 0.954 & 0.357 \\ \hline
$\eta=1$ & 0.990 & 0.974 & 0.849 & 0.695 &  - \\ \hline
\end{tabular}
\caption{PS detection accuracy $\rho_t$ as a function of sparsity and noise level $\eta$. The quality of the analysis was too poor to compute trajectories when both sparsity and noise were maximal.}
\label{tab:tmatch}
\end{table}

\begin{table}[t]
\begin{tabular}{*{6}{|c}|}
\hline & $256 \times 256$ & $64 \times 64$ & $32 \times 32$ & $16 \times 16$ & $8 \times 8$\\ \hline
$\eta=0$ & 1.1 & 1.1 & 1.4 & 4.8 & 9.9\\ \hline
$\eta=0.1$ & 1.2 & 1.3 & 1.6 & 4.6 & 9.6 \\ \hline
\;$\eta=0.3$\;  & 1.7 & 1.8 & 2.3 & 4.9 & 9.4 \\ \hline
$\eta=1$ & 2.3 & 3.1 & 4.5 & 6.9 & - \\ \hline
\end{tabular}
\caption{PS location precision $\rho_d$ (in units of the fine mesh) as a function of sparsity and noise level $\eta$.}
\label{tab:dmatch}
\end{table}

Tables \ref{tab:tmatch} and \ref{tab:dmatch} quantify the accuracy and precision of the algorithm in the presence of both noise and sparsity.
In the following discussion, we will use the benchmark analysis as the reference.
For each level of noise and sparsity, we compare the computed trajectories with the reference trajectories as follows. 
At each frame, every PS is matched with the nearest reference PS of the same chirality, provided that their separation is no greater than a fixed fraction $\alpha$ of the mean PS separation $L$. 
Each PS trajectory is then paired with all reference PS trajectories with which it was matched for at least a given fraction $\gamma$ of the period $T$, with all other matches discarded.
(We allow for a trajectory to be matched to multiple other trajectories to account for the possibility that some trajectories might be broken up by short gaps.)
Finally, we compute $\rho_d$, the average distance between the detected and reference PS across all matches, as well as the ratio
\begin{align}
\rho_t = \frac{2\sum_t m_t}{\sum_t (r_t + n_t)}
\end{align}
where the sum is over all frames and $m_t$, $r_t$, and $n_t$ are the number of matches, reference PSs, and detected PSs, respectively, in frame $t$;
$\rho_t$ ranges from 0 to 1 and represents the PS detection accuracy.
The results in the tables summarize the results for all data sets using fairly strict parameter choices $\alpha = 0.35$ and $\gamma = 0.8$.
(For our surrogate data, this means two trajectories are matched only if their separation is at most 16 units for at least 43 frames.)

The performance of the algorithm was quite good in a wide range of conditions.
Note that the accuracy and precision is imperfect even in the case of full resolution and no noise;
this is because of the temporal smoothing used in this analysis, which is not present in the benchmark.
PS detection accuracy is above 98\%, with precision of about 5\% of the mean PS separation or better, for 10 out of the 20 data sets, even in one case when $\eta=1$.
(When the accuracy is close to 100\%, most of the mismatches are due to PS pair creation or annihilation events being detected a few frames earlier or later than in the reference analysis, an error that has little to no dynamical importance.)
In most of the $16 \times 16$ data sets, for which the coarse grid resolution is approximately equal to $\alpha L$, the accuracy remains above 95\% and the precision is better than 5 spatial units (i.e., about 1 mm), demonstrating that the algorithm can locate PSs with sub-grid precision. 
The analysis only breaks down for data sampled on $8 \times 8$ grids, which is near the theoretical limit of such a method as the grid spacing (32 units) is comparable to the mean PS separation (46 units). The issue of maximal sparsity has been considered previously\cite{rappel2013} for a single spiral on a domain larger than the wavelength. Unfortunately, the results of that study cannot be directly compared to our case, where the average spacing between PSs is smaller than the wavelength.

\subsection{Comparison with alternative approaches}

To validate our algorithm, we also implemented and tested its most robust alternative, the Jacobian determinant method \cite{li2018}. 
The method essentially identifies PSs with the extrema of the two-dimensional field
\begin{align}
\mathcal{D}(t)=
\left|\begin{matrix}
\frac{\partial u}{\partial x}\big|_t & \frac{\partial u}{\partial y}\big|_t\\
\frac{\partial u}{\partial x}\big|_{t+\tau} & \frac{\partial u}{\partial y}\big|_{t+\tau}
\end{matrix}\right|
=\hat{z}\cdot(\nabla u|_t\times\nabla u|_{t+\tau}),
\end{align}
where $\tau$ is an empirically chosen time delay.
This method was used to compute PS trajectories for the full resolution ($256 \times 256$) data sets with varying levels of noise, and the results were compared to our benchmark analysis.
The time delay parameter was chosen to be $\tau=0.1125T = 6$ frames, which is approximately the value found to work best in the original paper.
Like our method, the Jacobian determinant method is better suited than traditional techniques to the presence of non-stationary PSs, noise, and/or sparsity.
However, we observed some difficulties not present in our approach.

First, the extrema of the field $\mathcal{D}$ must be defined with care as the less pronounced peaks do not correspond to PSs. 
As a result, some minimum peak prominence must be selected on a case-by-case basis for each recording and it is unclear how to determine the optimal choice without a refence.
Such an approach may not be feasible for experimental recordings, especially in the presence of spatial heterogeneity, which might cause the optimal threshold to vary in space.
For our data, we obtained the highest accuracy when labeling as PSs all extrema with a prominence exceeding the maximum value of $|\mathcal{D}|/3$.
Second, even in the absence of noise, some peaks corresponding to a single non-stationary PS were found to separate into multiple local extrema of similar magnitude, making it difficult to determine the exact number of PSs and their locations.
Finally, spatial smoothing of $\mathcal{D}$ was necessary to eliminate spurious PSs for $\eta$ as small as 0.1 (the lowest noise level we tested).
This step substantially reduces precision of PS location when the data are both noisy and sparse.

We smoothed $\mathcal{D}$ using a Gaussian kernel with a width of 2 grid spacings, which is in close correspondence with the authors' suggested approach.
In fact, this substantially improved accuracy even in the absence of noise, likely because of the second difficulty mentioned previously.
The results of the comparison of the smoothed Jacobian determinant method with the benchmark are summarized in Table \ref{tab:jd}.
The Jacobian determinant method performed very well for $\eta\leq 0.1$, achieving over 95\% accuracy, but accuracy fell below 90\% at $\eta=0.3$, and for $\eta=1$ no meaningful results could be produced as some frames contained over 100 spurious PSs.
This comparison demonstrates that the Jacobian determinant method can handle moderate levels of noise but falls apart at the higher noise levels that can be easily handled by our algorithm.
Our approach also handles sparse data better, locating PSs with sub-grid resolution.

\begin{table}[t]
\begin{tabular}{*{5}{|c}|}
\hline & $\eta=0$ & $\eta=0.1$ & $\eta=0.3$ & $\eta=1$ \\ \hline
$\;\rho_t\;$ & \;0.969\; & 0.956 & 0.895 & - \\ \hline
$\rho_d$ & 1.7 & 1.7 & 1.7 & - \\ \hline
\end{tabular}
\caption{PS detection accuracy $\rho_t$ and precision $\rho_d$ for the Jacobian determinant method as a function of noise level $\eta$. Analysis quality was too poor to compute trajectories for $\eta=1$.}
\label{tab:jd}
\end{table}

\section{Discussion}
\label{sec:discussion}

We illustrated a robust approach to rotor mapping using measurements of the transmembrane voltage $u$ obtained using a highly simplified model of atrial fibrillation. Furthermore, we used a rather unconventional approach where the positions of the rotors are defined using the intersections of the zero level sets of $\dot{u}$ and $\ddot{u}$.
One therefore might question the applicability of our results to electrophysiology studies using basket catheters, in which unipolar electrodes produce signals with a shape that is very different from the shape of the underlying transmembrane voltage.

Let us start by addressing the temporal profile of the voltage signal. Our particular choice of level sets was motivated by their relation to the excited and refractory phases of cardiomyocyte dynamics. Specifically, $\partial E$ describes the wavefront and waveback, while $\partial R$ describes the leading and trailing edges of the refractory region. This relationship makes it easy to use the topological and geometrical information encoded by the level sets to describe the dynamical mechanisms responsible for initiation, maintenance, and termination of cardiac arrhythmias. It also enables automatic classification of the topologically distinct events leading to an increase/decrease in the number of PSs and the associated increase/decrease in the complexity of the excitation pattern \cite{marcottetop}.

However, our particular choice of the two level sets is neither unique nor necessarily the best. For instance, in the presence of several inflection points in the repolarization phase of the action potential, which is typical for atrial tissue, these level sets might define additional spurious ``wavefronts'' and ``wavebacks.'' As an alternative,
one could use the zero level sets of $\dot{u}$ and $\dot{v}$, where $v$ is the gating variable \cite{marcottetop}, or the level sets $\phi\ {\rm mod}\ \pi=0$ and  $\phi\ {\rm mod}\ \pi=\pi/2$ of the phase field $\phi$ reconstructed using certain features of the voltage trace \cite{gurevich2017level}. For stationary (i.e., nondrifting, nonmeandering) spiral waves, the definitions of PSs based on these different choices are equivalent. The difference only becomes noticeable for the quickly moving rotors that are not targets of ablation therapy. A more common choice\cite{BaKnTu90} is to define one of the level sets using the transmembrane voltage itself, $u=u_{th}$. This choice defines the location of a spiral tip rather than a PS, and leads to a decrease in the accuracy of localization: unlike the PS, which is stationary, the spiral tip circles the PS for a stationary spiral wave.
 
Unipolar electrograms generated by basket catheters tend to have multiple maxima and minima per cycle \cite{kuklik2015} and do not provide a direct interpretation in terms of the depolarization/repolarization phase. They, however, have a characteristic feature (pronounced minima of the derivative $\dot{V}$ of the voltage $V$) that can be used to define the phase $\phi$ as a piecewise linear continuous function \cite{gurevich2017level}. Alternatively, the phase can be obtained using a temporal Hilbert transform. In either case, as long as the phase field $\phi(x_i,y_j,t_n)$ on a discrete regular grid can be reliably determined, our method can be applied rather directly \cite{gurevich2017level} to identify and track PSs using different level sets of $\phi$.

Another limitation of our study is that Gaussian noise is not representative of noise encountered in
clinical voltage mapping studies. One of the major sources of signal distortions is motion artifacts caused by
movement of the myocardial wall and/or poor contact of the electrode with the myocardial surface. Another major source of distortion is the far field effects, i.e., contamination of atrial electrograms by the much stronger electrical signal generated by the ventricles. Far field distortions can be mostly eliminated using, e.g., a single beat cancellation method \cite{zeemering2012}.
 
Finally, let us point out that we chose to use simple finite differences to compute temporal derivatives of the voltage due to the simplicity of this method. However, other methods could be used to further improve robustness of the algorithm for noisy data. The total variation regularized derivative \cite{Rudin1992} is one extremely robust option, although it is numerically costly. Another alternative with less computational overhead is least-squares polynomial interpolation \cite{knowles2014}. We have not pursued these more complicated methods, since even simple finite differencing produces rather impressive results. \\

\section{Conclusions}
\label{sec:conclusions}

We have introduced a novel approach for identifying the locations of multiple phase singularities associated with complicated patterns of excitation waves based on measurements of a single scalar field.
Because of its global nature, the new method was found to be substantially more robust than any previous, essentially local, methods aimed at identifying ``organizing centers'' of spiral wave activity.
In particular, we have demonstrated that our approach can simultaneously identify and locate tens of phase singularities, including ones that are highly nonstationary.
Moreover, their locations can be tracked in time with subgrid precision for data that are both very noisy (with noise level exceeding the signal level) and very sparse (on grids with spacing comparable to the mean separation between phase singularities).
This enables collection of a wide range of statistical information that can be used in model validation, for instance, and has a potential to impact applications such as clinical electrophysiology studies using intra-cardiac multi-electrode basket catheters.
It is worth emphasizing, however, that this method is not restricted to cardiac tissue and can be applied to any two-dimensional excitable system. 
Extensions to three dimensions are possible as well but are outside of the scope of this paper.

\section*{Supplementary material}
A movie showing the dynamics of PSs and the level sets corresponding to the wavefronts, wavebacks, and the leading/trailing edges of the refractory region in the benchmark analysis is provided as supplementary material.

\begin{acknowledgments}
DG gratefully acknowledges the support of the GT College of Sciences Undergraduate Research Science Award. This research was supported in part by NSF Grant No. PHY17-48958, NIH Grant No. R25GM067110, and the Gordon and Betty Moore Foundation Grant No. 2919.01.
\end{acknowledgments}

\section{References}
\input{Chaos2018.bbl} 

\appendix*

\section{Parameter sets}
For the reference analysis, the parameters used were $MPP_1 = 0.01, MPP_2=0.05, \delta=0.2, \sigma_s = \sigma_t = 0$ (no initial smoothing), and $\sigma_d = 4$. 
For the noisy/sparse data, $\sigma_t$ was changed to 5 and the minimum peak prominences depended on the noise as indicated in Table \ref{tab:mpp}.

\begin{table}[b]
\begin{tabular}{*{5}{|c}|}
\hline & $\eta=0$ & $\eta=0.1$ & $\eta=0.3$ & $\eta=1$ \\ \hline
$\;MPP_1\;$ & \;0.01\; & 0.05 & 0.25 & 0.25 \\ \hline
$MPP_2$ & 0.05 & 0.1 & 0.25 & 0.25 \\ \hline
\end{tabular}
\caption{Minimum peak prominences used for different levels of noise $\eta$.}
\label{tab:mpp}
\end{table}

\end{document}

%% file: Chaos2018.bbl
%